\documentclass[aps,prb,twocolumn]{revtex4}
\usepackage{graphicx}

\begin{document}


\title{Pressure-induced unconventional superconductivity in the  heavy-fermion antiferromagnet CeIn$_3$: An $^{115}$In-NQR study under pressure}

\author{S.~Kawasaki$^{1,*}$}%
\author{M.~Yashima$^1$}
\author{Y.~Kitaoka$^1$}
\author{K.~Takeda$^2$}
\author{K.~Shimizu$^3$}
\author{Y.~Oishi$^4$}
\author{M.~Takata$^4$}
\author{T.~C.~Kobayashi$^5$ }
\author{H.~Harima$^6$}
\author{S.~Araki$^7$}
\author{H.~Shishido$^{7,**}$}%
\author{R.~Settai$^7$}
\author{Y.~\=Onuki$^7$}

\affiliation{$^1$Department of Materials Engineering Science, Graduate School of Engineering Science, Osaka University, Toyonaka, Osaka 560-8531, Japan\\$^2$Department of Electrical and Electronic Engineering, Muroran Institute of Technology, Mizumoto, Muroran, 050-8585, Japan\\$^3$KYOKUGEN, Research Center for Materials Science at Extreme Conditions, Osaka University, Toyonaka, Osaka 560-8531, Japan\\ $^4$Japan Synchrotron Radiation Research Institute, Sayo, Hyogo 679-5198, Japan\\$^5$Department of Physics, Faculty of Science, Okayama University, Okayama 700-8530, Japan\\$^6$Department of Physics, Kobe University, Nada, Kobe 657-8501, Japan\\$^7$Department of Physics, Graduate School of Science, Osaka University, Toyonaka, Osaka 560-0043, Japan}%

\email[]{kawasaki@science.okayama-u.ac.jp}

\date{\today}

\begin{abstract}

We report on the pressure-induced unconventional superconductivity (SC) in the heavy-fermion (HF) antiferromagnet CeIn$_3$ by means of nuclear-quadrupole-resonance (NQR) studies conducted under a high pressure. The temperature ($T$) and pressure ($P$) dependences of the In-NQR spectra have revealed a first-order quantum-phase transition (QPT) from an antiferromagnetism (AFM) to paramagnetism (PM) at a critical pressure $P_{\rm c}=2.46$ GPa at which AFM disappears with a minimum value of $T_{\rm N}(P_{\rm c})$ = 1.2 K. High-energy X-ray scattering measurements under $P$ show a progressive decrease in the lattice density without any change in the crystal structure, whereas an increase in the NQR frequency ($\nu_{\rm Q}$) indicates an increase in the hybridization between $4f$ electrons and conduction electrons, which stabilizes the HF-PM state. This competition between the AFM phase where $T_{\rm N}$ is reduced and the formation of the HF-PM phase triggers the first-order QPT at $P_{\rm c} = 2.46$ GPa. Despite the lack of an AFM quantum critical point in the $P-T$ phase diagram, we highlight the fact that the unconventional SC occurs in both phases of the AFM and PM. The measurements of the nuclear spin-lattice relaxation rate $1/T_1$ in the AFM phase have provided evidence for the uniformly coexisting AFM+SC phase. Remarkably, the significant increase in $1/T_1$ upon cooling in the AFM phase has revealed the development of low-lying magnetic excitations down to $T_{\rm c}$ in the AFM phase; it is indeed relevant to the onset of the uniformly coexisting AFM+SC phase. In the HF-PM phase where AFM fluctuations are not developed, $1/T_1$ decreases without the coherence peak just below $T_{\rm c}$, followed by a power-law like $T$ dependence that indicates an unconventional SC with a line-node gap. Remarkably, $T_{\rm c}$ has a peak around $P_{\rm c}$ in the HF-PM phase as well as in the AFM phase. In other words, an SC dome exists with a maximum value of $T_{\rm c} = 230$ mK around $P_{\rm c}$, indicating that the origin of the pressure-induced HF SC in CeIn$_3$ is {\it not relevant to AFM spin fluctuations but to the emergence of the first-order QPT} in CeIn$_3$. These novel phenomena observed in CeIn$_3$ should be understood in terms of the first-order QPT because these new phases of matter are induced by applying $P$.  When the AFM critical temperature is suppressed at the termination point of the first-order QPT, $P_{\rm c} = 2.46$ GPa, the diverging AFM spin-density fluctuations emerge at the critical point from the AFM to PM. The results with CeIn$_3$ leading to a new type of quantum criticality deserve further theoretical investigations.

\end{abstract}


\maketitle 
\section{introduction}

In $f$-electrons based compounds, the hybridization between $f$ electrons and conduction electrons results in interesting physical phenomenon via the formation of a heavy-fermion (HF) state at low temperatures \cite{KitaokaKuramoto}. In particular, since the discovery of the first HF superconductor CeCu$_2$Si$_2$ in 1979 \cite{Steglich}, the HF superconductivity (SC) has attracted remarkable attentions as a candidate for understanding the novel mechanism for the unconventional SC discovered in the strongly correlated electron  systems (SCES) \cite{Kitaoka}. A common type of SC is based on bound electron pairs coupled via the lattice vibration \cite{BCS}. However, the SC in SCES including many HFs, cuprates and organic superconductors appears to have another binding force that forms Cooper pairs via electron-electron correlation. In particular, a number of studies on $f$-electron compounds revealed that an unconventional SC arises at or close to a second-order quantum-phase transition (QPT), i.e., the quantum critical point (QCP), where the magnetic order disappears at $T$ = 0 as a function of lattice density due to the application of hydrostatic pressure ($P$). In other words, near the magnetic order, the magnetic interaction between electron spins can mediate attractive interactions between the charge carriers. Phase diagrams have been obtained for antiferromagnetic HF compounds such as CePd$_2$Si$_2$ \cite{Grosche,Mathur,Grosche2}, CeIn$_3$ \cite{Walker,Mathur,Grosche2,Muramatsu2,Knebel,ShinjiJPSJ}, and CeRh$_2$Si$_2$ \cite{Movshovic,Araki}; these are schematically shown in Fig.~1(a). Significantly different behavior, schematically shown in Fig.~1(b), has been observed in the archetypal HF superconductor CeCu$_2$Si$_2$ \cite{Steglich,Bellarbi,Bellarbi2,Ykawasaki,Ykawasaki2} and the more recently discovered CeRhIn$_5$ \cite{Hegger,Muramatsu,Yashima}. Although both compounds have demonstrated an analogous behavior relevant to a magnetic QCP, it is noteworthy that an associated superconducting region extends to higher densities than in other compounds; their $T_{\rm c}$ value reaches its maximum away from point at which antiferromagnetism (AFM) is achieved \cite{Bellarbi,Bellarbi2,Muramatsu}. Most interestingly, the previous nuclear-quadrupole-resonance (NQR) studies have revealed that the AFM and SC coexist microscopically and that the SC does not exhibit any trace of a line-node gap opening in the low-lying excitations below $T_{\rm c}$ that are characteristic of the HF superconductors reported thus far \cite{Ykawasaki2,Mito,Skawasaki}.  

\begin{figure}[htbp]
\centering
\includegraphics[width=7cm]{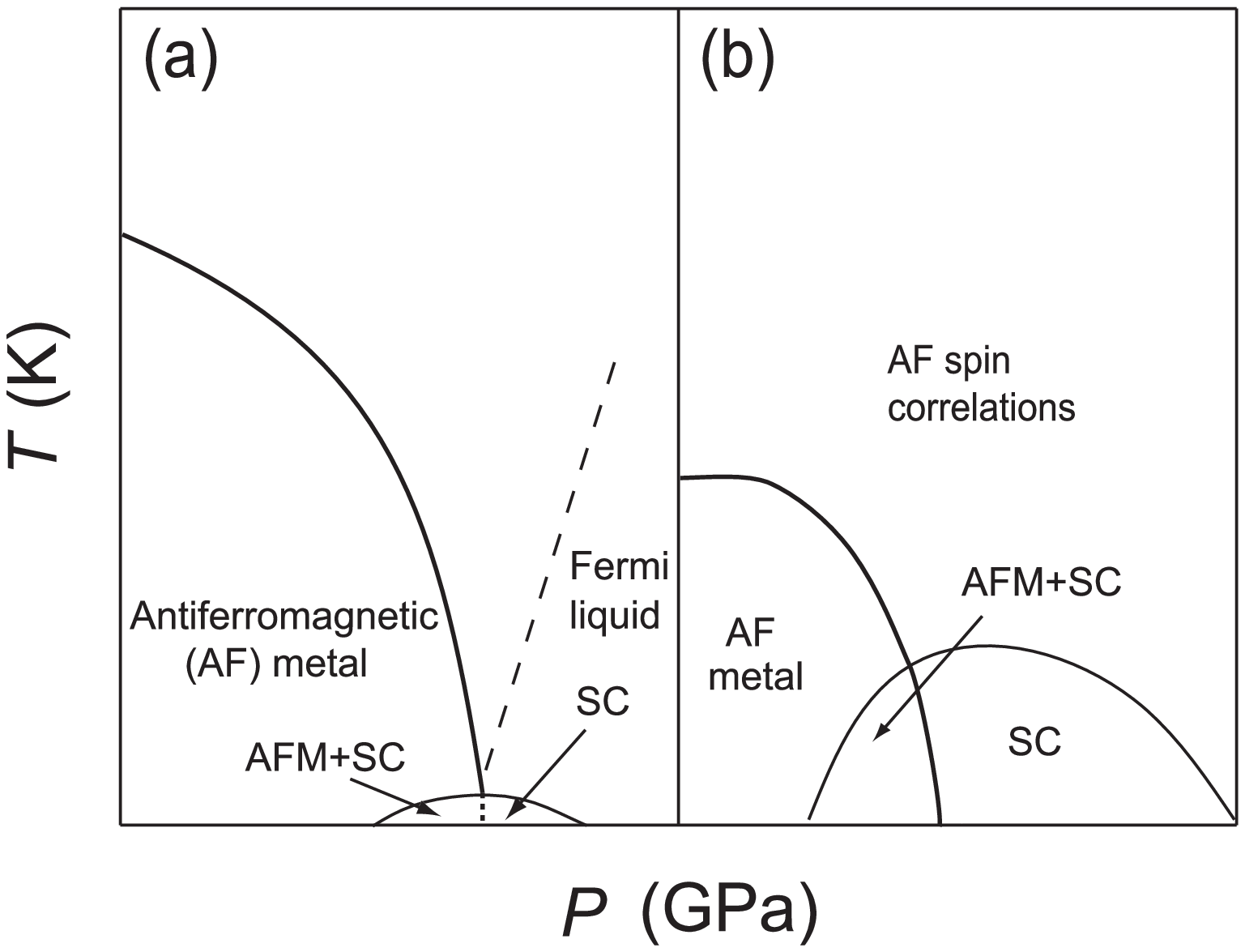}
\caption[]{Schematic phase diagrams of Ce-based heavy-fermion compounds: (a) for CePd$_2$Si$_2$\cite{Grosche,Mathur,Grosche2}, CeIn$_3$\cite{Walker,Mathur,Grosche2,Muramatsu2,Knebel} and CeRh$_2$Si$_2$\cite{Movshovic,Araki}; dashed line indicates the crossover; (b) for CeCu$_2$Si$_2$\cite{Steglich,Bellarbi,Bellarbi2,Ykawasaki,Ykawasaki2} and CeRhIn$_5$\cite{Hegger,Muramatsu,Yashima}.}
\label{fig1}
\end{figure}

Recently, it has been demonstrated by an extensive NQR study conducted under a high pressure that the novel $P-T$ phase diagram at zero magnetic field ($H$ = 0) in CeRhIn$_5$ is characterized by a tetra-critical point separating the pure AFM phase, the uniformly coexisting phase of AFM+SC, the SC phase, and the paramagnetic (PM) phase. Note that the AFM phase transition occurs inside the SC below $T_{\rm c}$ when a tetra-critical point is exceeded \cite{Yashima}. This result has revealed a close relationship between AFM and SC---both phases may be mediated by the same magnetic interaction. 
In contrast, the two superconducting domes have been reported to be a function of $P$ in the case of CeCu$_2$(Si$_{1-x}$Ge$_{x}$)$_2$ \cite{Yuan}; however, the origin of the SC in HF compounds is still a topical issue. One dome (SC1) is formed around the AFM QCP, whereas the other (SC2) emerges in the HF state without any indication of AFM spin fluctuations because the system is still far from the point of the AFM QCP. Interestingly, the maximum $T_{\rm c}$ value in SC2 as the function of $P$ is higher than that in SC1 in the case of CeCu$_2$(Si$_{1-x}$Ge$_{x}$)$_2$. Although the possible origins of SC2 are not yet known, a new type of pairing mechanism, besides AFM spin fluctuations, has been suggested for mediating the Cooper pairs in HF systems. For instance, valence fluctuations of Ce ions  may be responsible for the onset of SC2 via the increase in hybridization between the Ce-4$f$ electrons and conduction electrons \cite{Yuan,Onishi,Miyake,Watanabe}. Two SC domes have also been suggested in the case of CeRh$_{1-x}$Ir$_x$In$_5$ \cite{Pagliuso,Nicklas,Shinji,Shinji2}. These results suggest that there are still underlying issues that remain to describe rich phases of matter appeared in the antiferromagnetic HF systems under $P$.

As shown in Fig. 2, CeIn$_3$ is formed in the cubic AuCu$_3$ structure and orders antiferromagnetically below the N\'eel temperature $T_{\rm N}$ = 10.2 K at an ambient pressure ($P$= 0). It has an ordering vector ${\bf Q}$ = (1/2,1/2,1/2) \cite{Morin} and Ce magnetic moment $M_{\rm AFM}\sim 0.5\mu_{\rm B}$, which were determined by NQR measurements \cite{Kohori,Kohori2} and the neutron-diffraction experiment on the single crystals \cite{Knafo}, respectively. The resistivity measurements of CeIn$_3$ revealed the $P-T$ phase diagram of the AFM and SC---$T_{\rm N}$ decreases with increasing $P$, and on almost near the point where AFM is achieved, SC emerges in a narrow $P$ range of approximately 0.5 GPa, thereby exhibiting a maximum value of $T_{\rm c}\sim 200$ mK at around $P_{\rm c}$ = 2.5 GPa where AFM disappears \cite{Walker,Mathur,Grosche2,Muramatsu2,Knebel}. A non-Fermi-liquid behavior was suggested from the $T^{3/2}$ dependence of resistivity within the framework of spin fluctuations theory \cite{MoriyaUeda} in narrow $P$ and $T$ ranges around $P_{\rm c}$. Then, it was inferred that magnetic fluctuations can mediate spin-dependent attractive interactions between the charge carriers in CeIn$_3$ \cite{Walker,Mathur,Grosche2,Muramatsu2,Knebel}.

\begin{figure}[htbp]
\centering
\includegraphics[width=5.5cm]{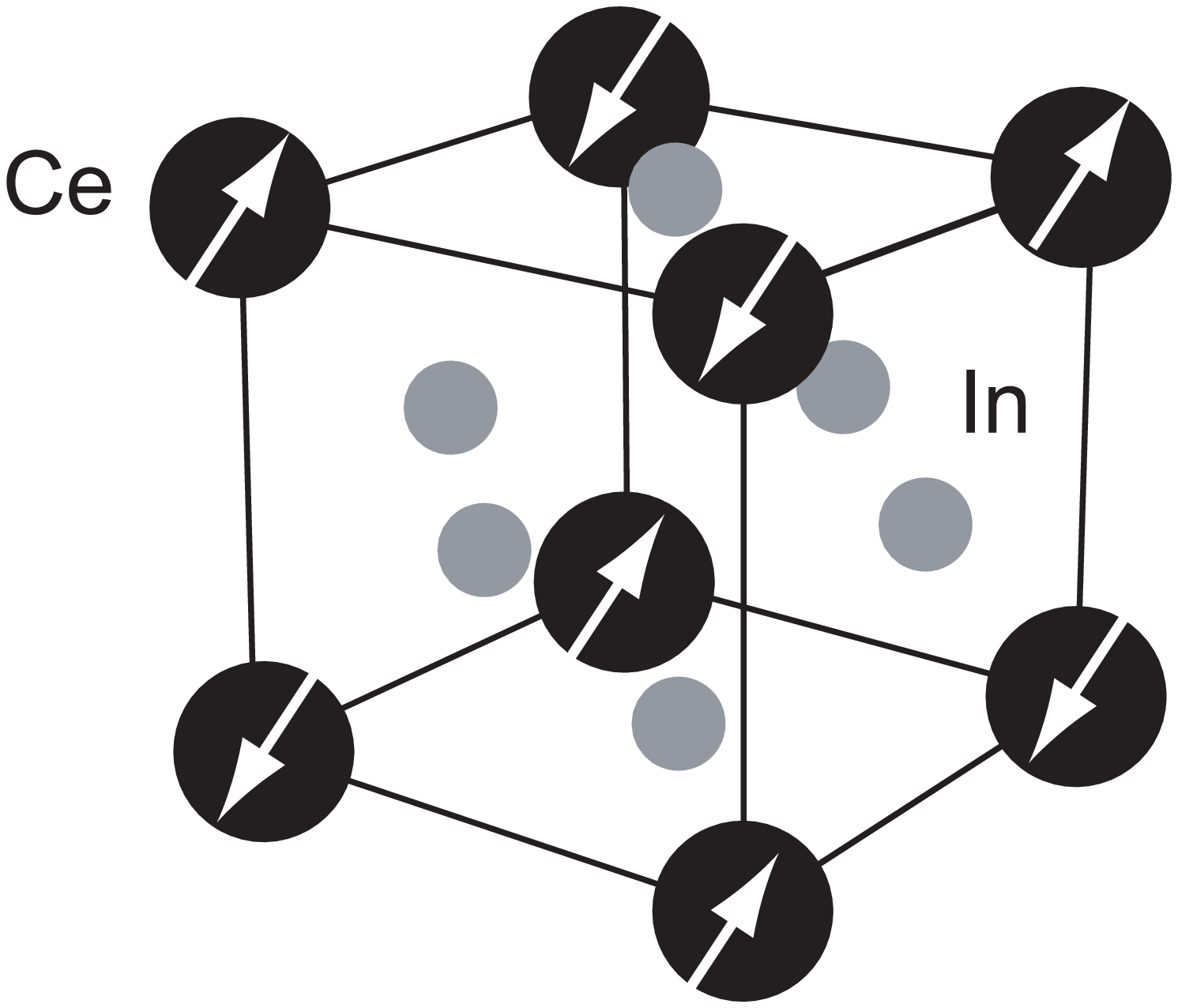}
\caption[]{Crystal and spin structures of CeIn$_3$ below $T_{\rm N}$.}
\label{fig2}
\end{figure}

 Previous NQR studies have revealed a systematic change in the magnetic character through the measurements of the nuclear-spin-lattice relaxation rate $1/T_1$ of the $^{115}$In-NQR under $P$ \cite{Thessieu,Skawasaki2,Skawasaki3,ShinjiJPSJ}. The localized magnetic character of 4$f$ magnetic moments is robust up to $P$ = 1.9 GPa. The characteristic temperature $T^{*}$, below which the system crosses over to an HF regime, increases dramatically with further increase in $P$.  As a result, the HF state becomes stable due to the increase in $T^{*}$ as $P$ increases beyond $P_{\rm c}$. The measurements of $1/T_1$ and ac-susceptibility ($\chi_{\rm ac}$) at $P$ = 2.65 GPa till $T$ =  50 mK provided the first evidence of an unconventional SC at $T_{\rm c}$ = 95 mK in CeIn$_3$, which arises in the fully established HF state below $T_{\rm FL}$ = 5 K \cite{Skawasaki3}. The phase separation into the AFM and PM phases in CeIn$_3$ is evidenced from the observation of two kinds of NQR spectra around $P_{\rm c}$ \cite{ShinjiJPSJ}. Nevertheless, it was demonstrated that the SC in CeIn$_3$ occurs in both phases at $P$ = 2.43 GPa, where the maximum value of $T_{\rm c}^{\rm max}$ =  230 mK is observed for the PM phase. Remarkably, the SC uniformly coexists with AFM below $T_{\rm c}$ = 190 mK \cite{ShinjiJPSJ}. 
Furthermore, a first-order QPT was suggested from a possible phase separation into the AFM and the HF-PM phases near $P_{\rm c}$. However, since the possibility that this phase separation near $P_{\rm c}$ is due to an inevitable distribution of $P$ inside the pressure cell cannot be ruled out, we cannot conclude whether QPT is of a first or a second order. In this paper, through the extensive In-NQR studies, we provide new insights into a novel $P-T$ phase diagram in CeIn$_3$ and into the origin of the unconventional SC that emerges in the vicinity of a first-order QPT transition from the AFM phase to the HF-PM phase.

\section{Experimental procedures}

High-quality single crystals of CeIn$_3$ were grown by the Czochralski method. They were moderately crushed into grains so that rf pulses can easily penetrate into the samples. However, in order to avoid crystal distortions, the grain diameters were kept larger than 100 $\mu$m. A small piece of CeIn$_3$ cut from the same batch as the sample used in the present work exhibited zero resistance in a range $P =  2.2 - 2.8$ GPa \cite{Muramatsu2}, which is in good agreement with the previous reports \cite{Mathur,Grosche2,Knebel}. An $^{115}$In-NQR spectrum was obtained by plotting the spin-echo intensity as a function of frequency. In order to detect an internal magnetic field associated with an onset of AFM around $P_{\rm c}$, the NQR spectrum for the 1$\nu_{Q}$ $(\pm 1/2\leftrightarrow \pm 3/2)$ transition was precisely obtained by the Fourier transform method of spin-echo signal. Under the condition that the NQR spectra result from both phases of AFM and PM in the vicinity of $P_{\rm c}$, each volume fraction was estimated from the NQR intensity $I(0)$ for the 1$\nu_{\rm Q}$ transition, which was precisely estimated through a fitting to $I(t)=I(0)\exp{(-t/T_2)}$, where $T_2$ is the nuclear spin-spin relaxation time. The $^{115}$In-NQR $T_1$ was measured by the conventional saturation-recovery method in a range of $T = 0.05 - 70$ K. The 2$\nu_{\rm Q}$ ($\pm 3/2\leftrightarrow \pm 5/2$) and 1$\nu_{\rm Q}$ ($\pm 1/2\leftrightarrow \pm 3/2$) transitions were used for the $T_1$ measurement above and below $T=1.4$ K, respectively. The high-frequency $\chi_{\rm ac}$ was measured by using an {\it in-situ} NQR coil \cite{Mito}. Hydrostatic pressure was applied by utilizing a NiCrAl-BeCu piston-cylinder type clamping cell filled with Si-based organic liquid as a pressure-transmitting medium \cite{Andrei}. In order to calibrate the pressure at low temperatures, the shift in $T_{\rm c}$ of the Sn metal under $P$ was measured by the conventional four-terminal resistivity measurement. To reach the lowest temperature of 50 mK, a $^3$He-$^4$He dilution refrigerator was used. 

The $^{115}$In-NQR spectra in the PM state at $T$ = 77 K and $P$ = 0 are shown in the top of Fig.~3(a) where four transitions are observed at different frequencies $\nu = n\nu_{\rm Q}$ for $n$ = 1, 2, 3 and 4. Here, $\nu_{\rm Q}$ is defined by the NQR Hamiltonian: $\mathcal{H}_{\rm Q}$ = $(h \nu_{\rm Q}/6)[3{I_z}^2-I(I+1)+\eta({I_x}^2-{I_y}^2)]$, where $\eta$ is the asymmetry parameter of the electric field gradient. Note that $\nu_{\rm Q}$ = 9.61 MHz and $\eta$ = 0 at $P$ = 0. Since the full width at the half maximum (FWHM) for the 1$\nu_{\rm Q}$-NQR spectrum is quite sharp at 60 kHz, the sample that is used here is confirmed to be of high quality. Under the condition that the AFM order sets in, the Hamiltonian of the $^{115}$In nuclei is replaced by $\mathcal{H}_{\rm AFM}=-\gamma\hbar\vec{I}\cdot\vec{H_{\rm int}} + \mathcal{H}_{\rm Q}$, where $\vec H_{\rm int}=(H_{\perp}, 0, H_{\parallel})$ is an internal magnetic field associated with the onset of the AFM order. In the case of CeIn$_3$, $H_{\parallel}$ is cancelled at the position of the In site. Hence, the onset of the AFM order for CeIn$_3$ is identified from the splitting of the 1$\nu_Q$ spectrum and the frequency shift of the other spectra due to the appearance of $H_{\perp}$.
\begin{figure}[htbp]
\centering
\includegraphics[width=6cm]{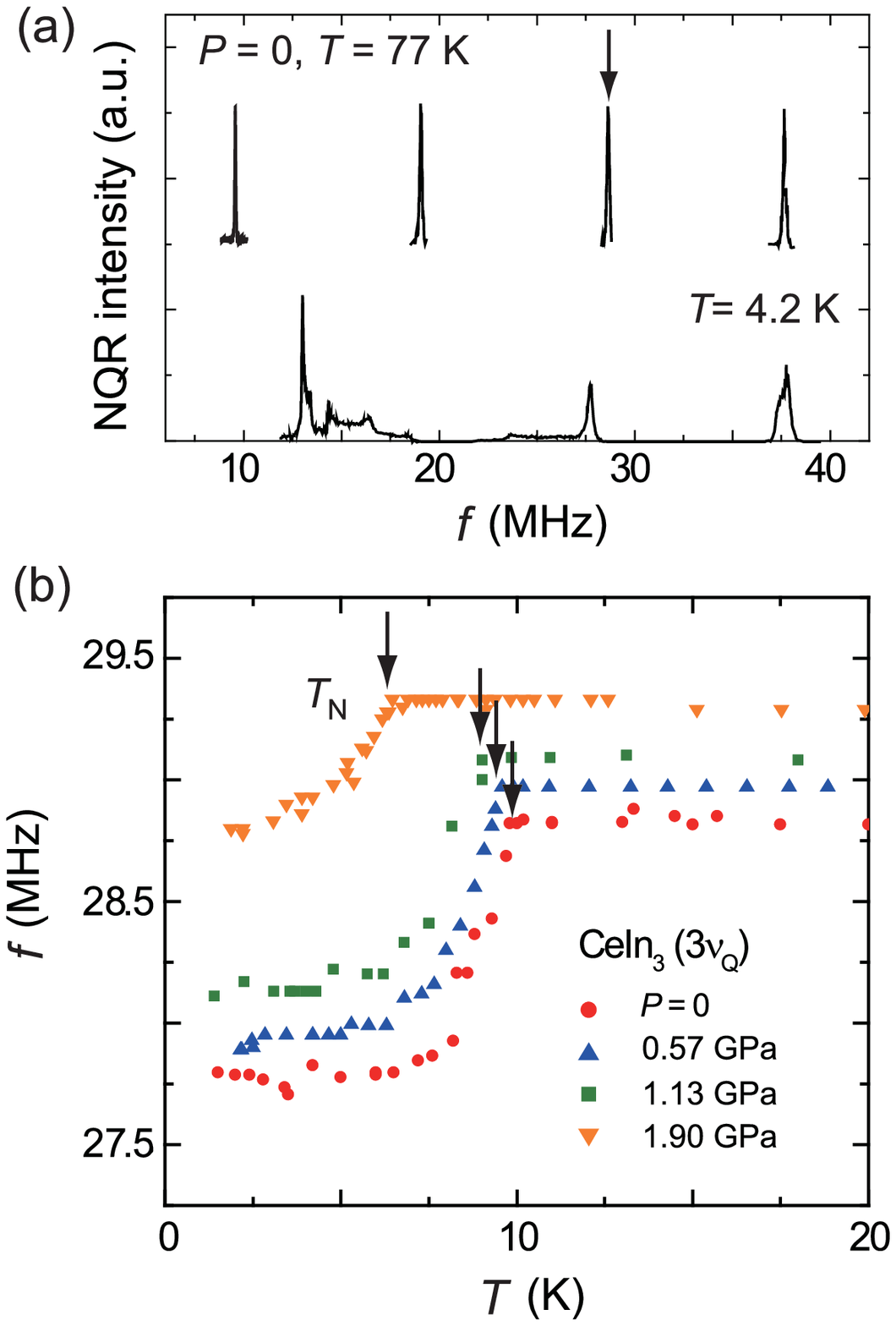}
\caption[]{(color online) (a) NQR spectra for CeIn$_3$ above and below $T_{\rm N}$ = 10.2 K at $P$ = 0.  Solid arrow indicates the position of the 3$\nu_{\rm Q}$ transition. (b) The $T$ dependence of the peak of the 3$\nu_{\rm Q}$ spectrum in a range of $P = 0 - 1.90$ GPa. The solid arrows point to $T_{\rm N}$. 
}
\label{fig3}
\end{figure}

In fact, the NQR spectra at $T$ = 4.2 K below $T_{\rm N}$ are indicated at the bottom of Fig.~\ref{fig3}(a), revealing a significant change due to the presence of $H_{\perp}$. Note that the 3$\nu_{\rm Q}$- and 4$\nu_{\rm Q}$- spectra remain extremely sharp even below $T_{\rm N}$, which guarantees that $H_{\rm int}$ is homogeneously determined at all the In sites below $T_{\rm N}$. These results are consistent with the previous NQR measurements \cite{Kohori,Kohori2}.
Extensive analyses of these spectra as the functions of $T$ and $P$ enable us to determine $T_{\rm N}(P)$ and the value of the antiferromgnetically ordered moment $M_{\rm AFM}(T,P)$ as functions of $P$ and $T$ for CeIn$_3$ even in close proximity to $P_{\rm c}$. 

Pressure dependence of the unit cell volume at low temperature was measured by powder X-ray diffraction experiments using synchrotron radiation in SPring-8 BL10XU. The experiment carried out by using a He-gas driven diamond-anvil cell. The pressure medium was methanol-ethanol mixture. The pressure was determined by the ruby fluorescence method. The lattice parameter were refined by the Rietveld method using the RIETAN-2000 program.\cite{Izumi}

\section{Evidence for pressure-induced first-order quantum phase transition from antiferromagnetism to paramagnetism}

\begin{figure}[htbp]
\centering
\includegraphics[width=6cm]{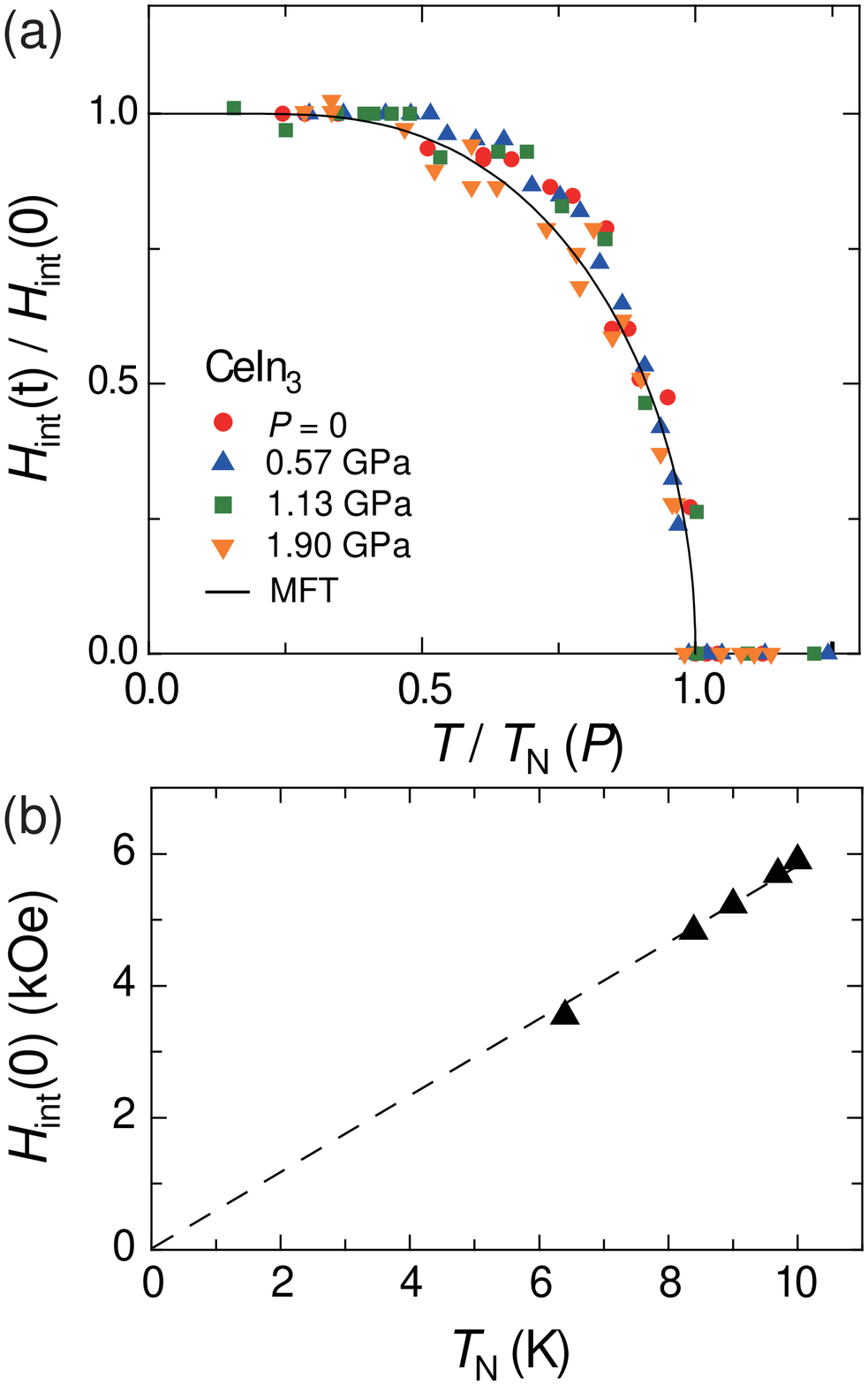}
\caption[]{(color online) (a) Plots of $H_{\rm int}(t)/H_{\rm int}(0)$ vs $T/T_{\rm N}(P)$ for AFM phase. Here, $H_{\rm int}(0)$ is a saturated value extrapolated to $T$ = 0. Solid curve indicates the molecular field theory (MFT) with $S$ = 1/2. (b) $H_{\rm int}(0)$ vs $T_{\rm N}$ plot in a $P$ range of $P = 0 - 1.90$ GPa. Dotted line is an eye-guide. 
}
\label{fig4}
\end{figure}

The $P$ dependences of $T_{\rm N}$ and $H_{\rm int}\propto M_{\rm AFM}$ in the AFM phase are deduced from the analysis of the 3$\nu_{\rm Q}$ spectrum (7/2$\Longleftrightarrow$5/2 transition) below $T_{\rm N}$, which is significantly shifted below $T_{\rm N}$, as denoted by the arrows in Fig.~\ref{fig3}(b). Plots of $H_{\rm int}(t)/H_{\rm int}(0)$ vs $t = T/T_{\rm N}(P)$ are presented for various $P$ in Fig.~\ref{fig4}(a). Here, $H_{\rm int}(0)$ is a value at $T$ = 0. Significantly, the $T$ dependence of $M_{\rm AFM}$ below $T_{\rm N}$ in a range of $P = 0 - 2$ GPa is in good agreement with the molecular-field theory (MFT) with spin $S$ = 1/2 and $m_{\rm AFM}$ = tanh($m_{\rm AFM}/t$) as indicated by the solid curve in Fig.~\ref{fig4}(a), reflecting a localized magnetic character of AFM in CeIn$_3$. It must be noted that in Fig.~\ref{fig4}(b), $H_{\rm int}\propto M_{\rm AFM}$ is in proportion to $T_{\rm N}$ till $P\sim$ 2 GPa. 

\begin{figure}[htbp]
\centering
\includegraphics[width=6.5cm]{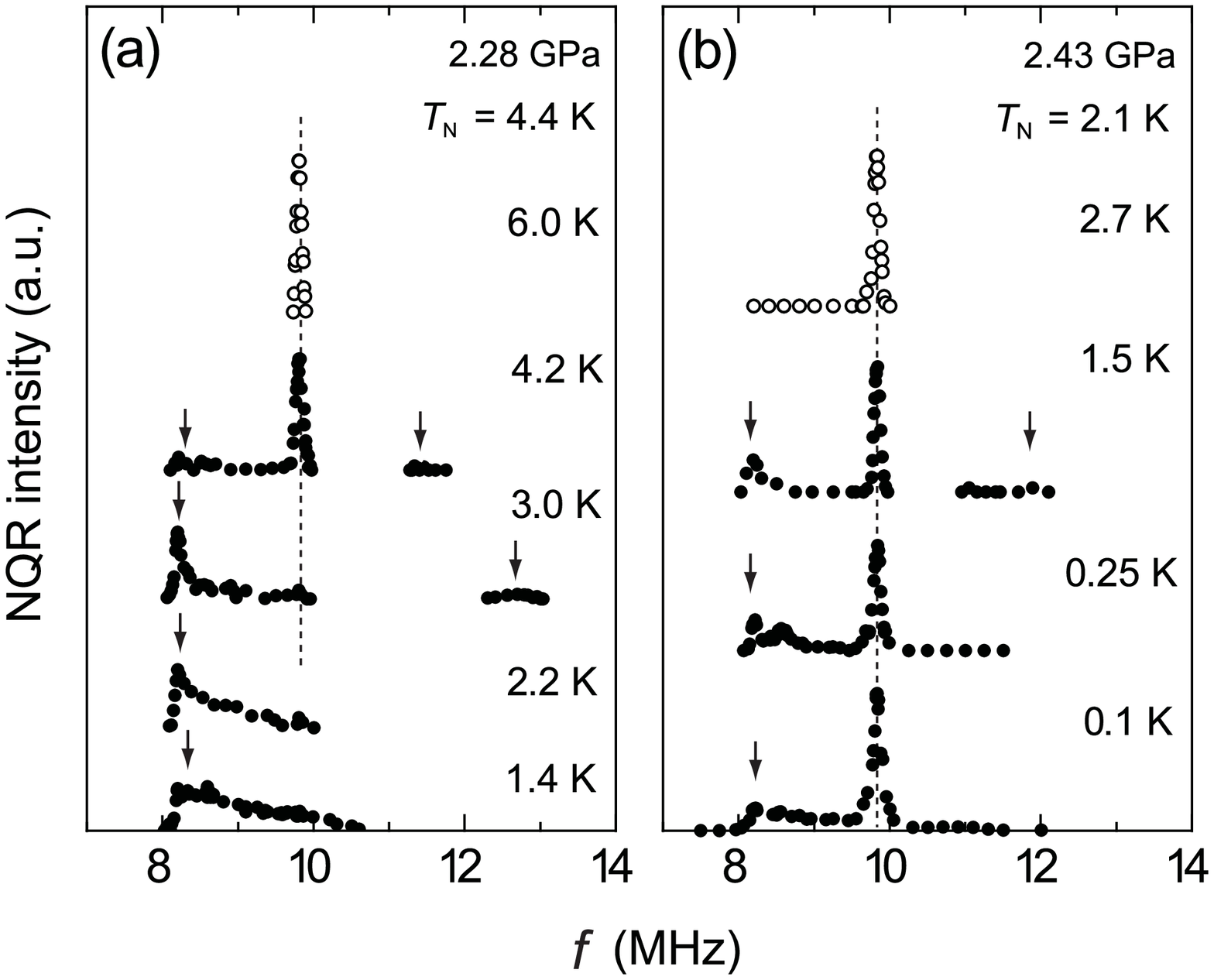}
\caption[]{Temperature dependence of 1$\nu_{\rm Q}$-NQR spectrum at (a) $P=$ 2.28 and (b) $P=$ 2.43 GPa just below $P_c=$ 2.46 GPa. Open and solid circles indicate the respective spectra above and below $T_{\rm N}$. The dotted line and solid arrows point to the respective frequencies where the peak in NQR spectrum is observed for the PM and AFM phases.}
\label{fig5}
\end{figure}

\begin{figure}[htbp]
\centering
\includegraphics[width=5.5cm]{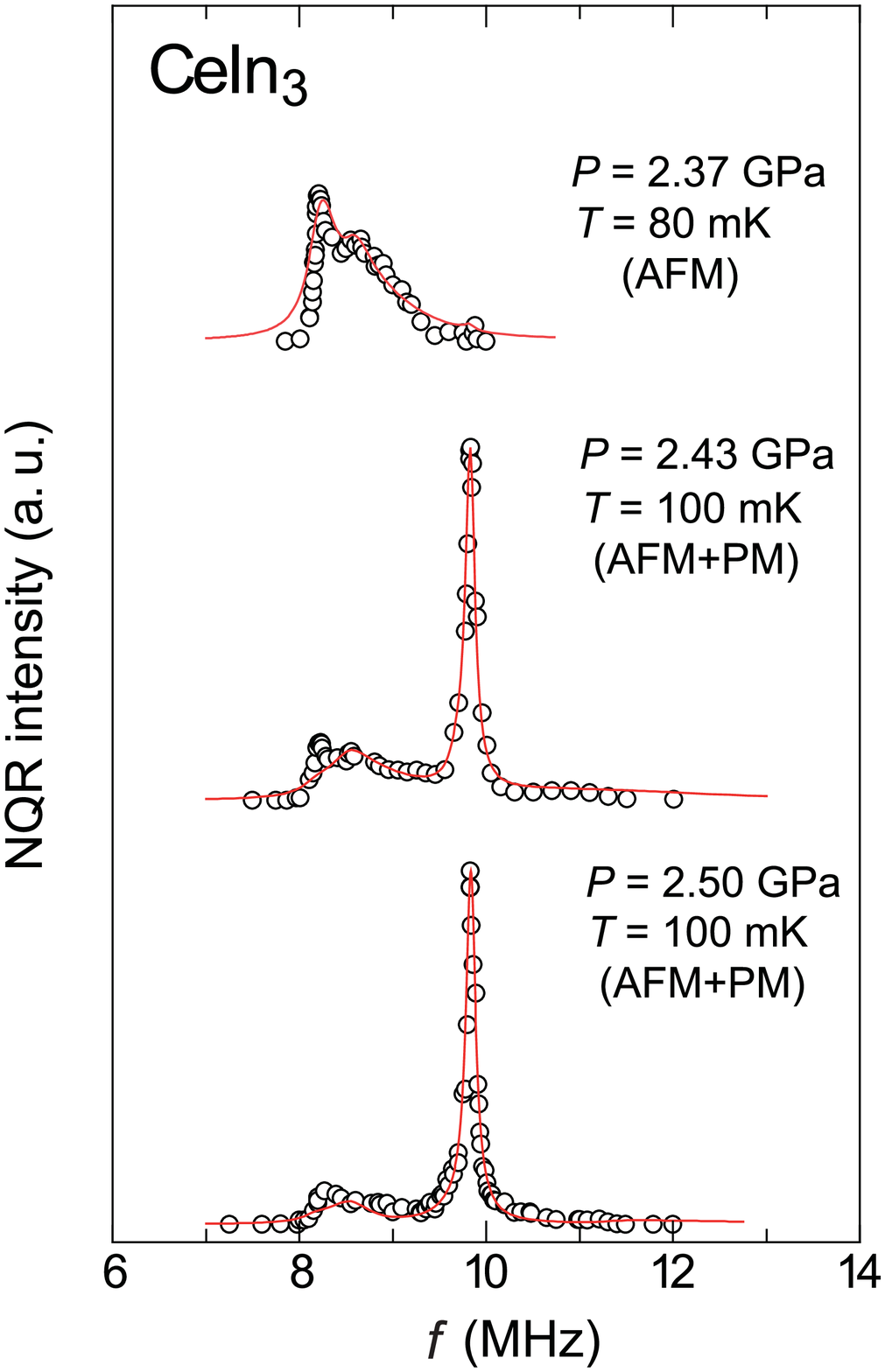}
\caption[]{(color online) 1$\nu_{\rm Q}$-NQR spectra for $P$ =  2.37 and 2.43 GPa just below $P_{\rm c}$ = 2.46 GPa and for $P$ = 2.50 GPa just above $P_{\rm c}$. Solid curves are the simulations assuming an inevitable $P$ distribution inside the sample in the pressure cell (see text). In the vicinity of $P_{\rm c}$, NQR spectra result from the AFM and PM phases, which are separated by the first-order quantum phase transition. Here, the NQR spectral intensity is normalized by a peak intensity of the NQR spectrum of the PM phase.}
\label{fig6}
\end{figure}

Although $T_{\rm N}$ and $H_{\rm int}(0)$ steeply decrease as $P$ approaches $P_{\rm c}$ above 2 GPa, the $T$ dependence of the 1$\nu_{\rm Q}$-NQR spectrum allows us to detect precisely the presence of an AFM order even in the vicinity of $P_{\rm c}$. In fact, Figs.~\ref{fig5}(a) and \ref{fig5}(b) show the $T$ dependences of those at $P$ = 2.28 and 2.43 GPa where the respective spectra above and below $T_{\rm N}$ are indicated by open and solid symbols. It should be noted that the spectrum due to the PM phase is observed at 3 K and 0.1 K and even below $T_{\rm N}$ = 4.4 K and 2.1 K at $P$ = 2.28 GPa and 2.43 GPa, respectively. This means that a phase separation into AFM and PM occurs as $P$ approaches $P_{\rm c}$. This result is corroborated by the $P$ dependence of the spectra at temperatures lower than $T_{\rm N}$, which are shown in the top, middle, and bottom parts of Fig.\ref{fig6} for $P$ = 2.37, 2.43, and 2.50 GPa, respectively. Apparently, the phase separation into the AFM and PM phases occurs in the vicinity of $P_{\rm c}$.

\begin{figure}[htbp]
\centering
\includegraphics[width=6cm]{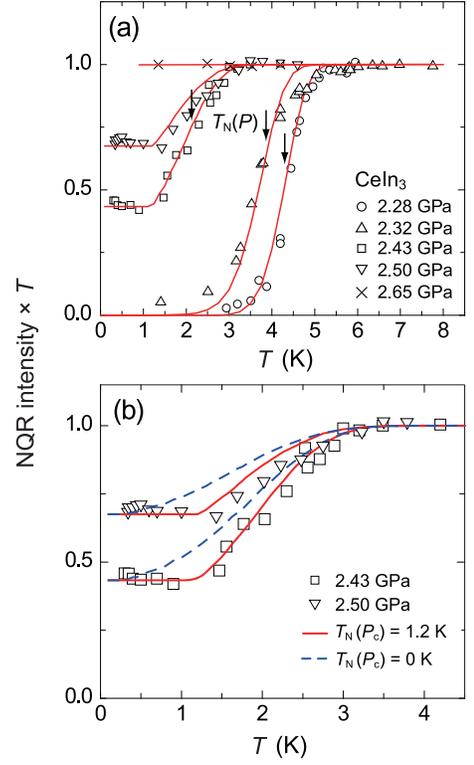}
\caption[]{(color online) (a) Temperature dependence of the NQR intensity $\times$ temperature for the 1$\nu_{\rm Q}$ transition around $P_{\rm c}$. Arrows indicate $T_{\rm N}$ at $P$ = 2.28, 2.32, and 2.43 GPa. Solid curve is a simulation assuming that $T_{\rm N}$ is distributed due to the inevitable $P$ distribution in the pressure cell (see text). (b) Temperature dependence of NQR intensity $\times$ temperature at $P$ = 2.43 and 2.50 GPa. Solid and dotted curves are simulations assuming first-order ($T_{\rm N}(P_{\rm c})$ = 1.2 K) and second-order ($T_{\rm N}(P_{\rm c})$ = 0 K) phase transitions, respectively.  }
\label{fig7}
\end{figure}

The NQR spectral intensity $I(T)$ increases with $1/T$ upon cooling, and $I(T)\times T$ is proportional to the number of observable In nuclei. Hence, the $I(T)\times T$ value for the NQR spectrum must be constant in the PM state. In fact, as shown in Fig.~\ref{fig7}, where the data are normalized by a value at high temperatures above $T_{\rm N}$ at each $P$, $I(T)\times T$ remains a constant for $P$ = 2.65 GPa larger than $P_{\rm c}$ where the AFM order collapses. In contrast, the $I(T)\times T$ value decreases to zero upon cooling for $P$ = 2.28 and 2.32 GPa, which are lower than $P_{\rm c}$. Unexpectedly, note that their decreasing behavior is not as steep below $T_{\rm N}$, suggesting a possible distribution of $T_{\rm N}$ due to an inevitable distribution of $P$ inside the sample. It is noteworthy that each $I(T)\times T$ for $P$ = 2.43 and 2.50 GPa near $P_{\rm c}$ decreases due to the onset of the AFM order below $T_{\rm N}$, but it becomes constant below $T$ $\sim$ 1.2 K in both pressures and remains a finite value at the lowest temperatures; this is corroborated by the NQR spectra at $T$ = 0.1 K as indicated in Fig.\ref{fig6}. These results reveal that both the AFM and PM phases are mixed at $P$ = 2.43 and 2.50 GPa near $P_{\rm c}$. Since $T_{\rm N}$ decreases steeply near $P_{\rm c}$, this mixture is associated with an inevitable distribution of $P$ inside the sample, revealing that the application of $P$ is not always homogeneous. 

\begin{figure}[htbp]
\centering
\includegraphics[width=6cm]{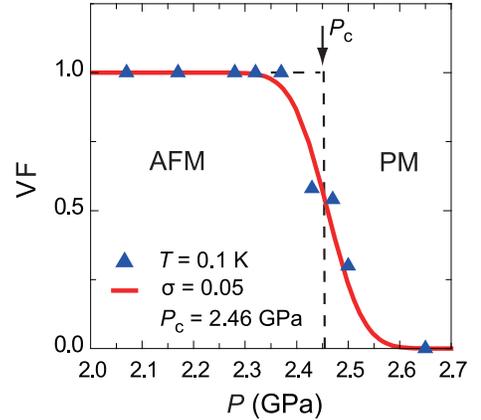}
\caption[]{(color online) Pressure ($P$) dependence of the volume fraction of the AFM phase at $T = 0.1$ K in the vicinity of $P_{\rm c}$. Solid curve is a simulation assuming a Gaussian-type $P$ distribution inside the sample in the $P$ cell (see text). Vertical dotted line indicates the first-order quantum phase transition for $\sigma = 0$ provided that a $P$ distribution is absent. Arrow points to  $P_{\rm c}$ = 2.46 GPa. }
\label{fig8}
\end{figure}

\begin{figure}[htbp]
\centering
\includegraphics[width=6cm]{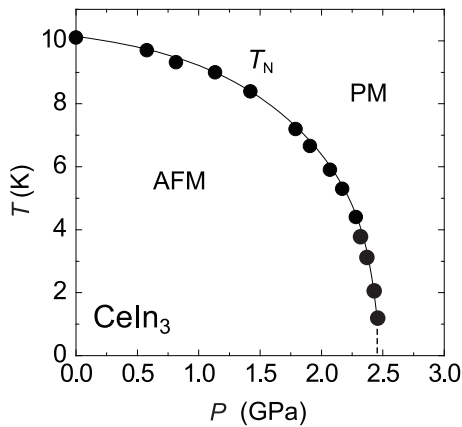}
\caption[]{Pressure dependence of $T_{\rm N}$ in CeIn$_3$. $T_{\rm N}$ above $P$ = 2.28 GPa are determined from the analysis of the $T$ and $P$ dependences of the NQR spectrum.  Solid curve is an eye-guide. Dotted line indicates $P_{\rm c}$ = 2.46 GPa where the first-order quantum phase transition separates AFM and PM.}
\label{fig9}
\end{figure}

Considering this experimental situation in mind, we attempt to reproduce the $P$ dependences of the NQR spectra in Fig. 6, temperature dependence of $I(T)\times T$ in various pressures in Fig. 7, and the volume fraction of the AFM ($V_{\rm AFM}$) in Fig. 8 by assuming the following Gaussian distribution of $P$ inside the sample: 

\begin{equation}
V(P_0,P) = \frac{1}{\sqrt{2\pi}\sigma}exp\left[-\frac{1}{2\sigma^2}(P-P_0)^2\right]
\end{equation}

and

\begin{equation}
\int^\infty_{-\infty} V(P_0,P) dP = 1
\end{equation}

Here, $P_{\rm 0}$ represents an external pressure, and $\sigma$ represents the mean deviation in the Gaussian distribution function. A $P$ dependence of $V_{\rm AFM}$ is obtained by the integration of $V(P_0,P)$ against $P$ from $-\infty$ to $P_{\rm c}$ as follow;

\begin{equation}
V_{\rm AFM}(P_0) = \int^{P_c}_{-\infty}V(P_0,P) dP
\end{equation}

Figure~\ref{fig8} shows the $P$ dependence of $V_{\rm AFM}$, which is determined by the NQR intensity at $T$ = 0.1 K (see Figs.~\ref{fig6} and \ref{fig7}). As shown in the solid curve in Fig.~\ref{fig8}, a best fit to the data is obtained with parameters $\sigma$ = 0.05 and $P_{\rm c}$ = 2.46 GPa. Here, a dotted line is drawn as a phase boundary at $T$ = 0 if a pressure distribution were absent. This {\it in-situ} $P$ distribution is comparable to the values in other experiments that were performed using a piston cylinder-type clamping cell \cite{Yu,Knebel2}. By using the above parameters of $\sigma$ = 0.05 and $P_{\rm c}$ = 2.46 GPa, we show that the spectra in the range of $P = 2.37 - 2.50$ GPa shown in Fig.~\ref{fig6} are consistently simulated by assuming that the $T$ dependence of $H_{\rm int}(T)$, which is induced by AFM moments, can be described in terms of the molecular-field model and the relation $T_{\rm N}\propto H_{\rm int}(0)$. Further, we assume the $P$ dependence of $T_{\rm N}$ as $T_{\rm N}(P)$ = 4.0$\times$$\left(\frac{2.475-P}{2.475-2.31}\right)^{0.5}$ just below $P_{\rm c}$ = 2.46 GPa.  As a result, we have obtained excellent fittings, as shown by the solid lines in Figs.~\ref{fig6} and \ref{fig7}. Figure~\ref{fig9} indicates the thus obtained $P$ dependence of $T_{\rm N}$ in CeIn$_3$. Remarkably, $T_{\rm N}$ disappears suddenly at a minimum value of $T_{\rm N} = 1.2$ K at $P_{\rm c}$ =  2.46 GPa, suggesting a weak first-order QPT from AFM to PM in CeIn$_3$ as the function of $P$. Figure 7(b) shows the $T$ dependences of $I(T)\times T$ at $P$ = 2.43 and 2.50 GPa and simulation curves assuming first-order phase transition with $T_{\rm N}(P_{\rm c})$ = 1.2 K (solid curves) and second-order phase transition with $T_{\rm N}(P_{\rm c})$ = 0 K (dotted curves), respectively. Notably, simulation for second-order phase transition does not fit experimental data at all. Especially, $I(T)\times T$ = constant behavior below $T_{\rm N}(P_{\rm c})$ = 1.2 K observed at $P$ = 2.43 and 2.50 GPa near $P_{\rm c}$ is a significant feature of first-order phase transition. This contrasts with the novel phase diagram of the HF antiferromagnet CeRhIn$_5$ under $P$ which is characterized by the tetra-critical point separating the pure AFM phase, the uniformly coexisting phase of AFM+SC, and the PM-SC phase.\cite{Yashima}

\section{Pressure-induced evolution of magnetic properties around $P_{\rm c}$}

\begin{figure}[h]
\centering
\includegraphics[width=7cm]{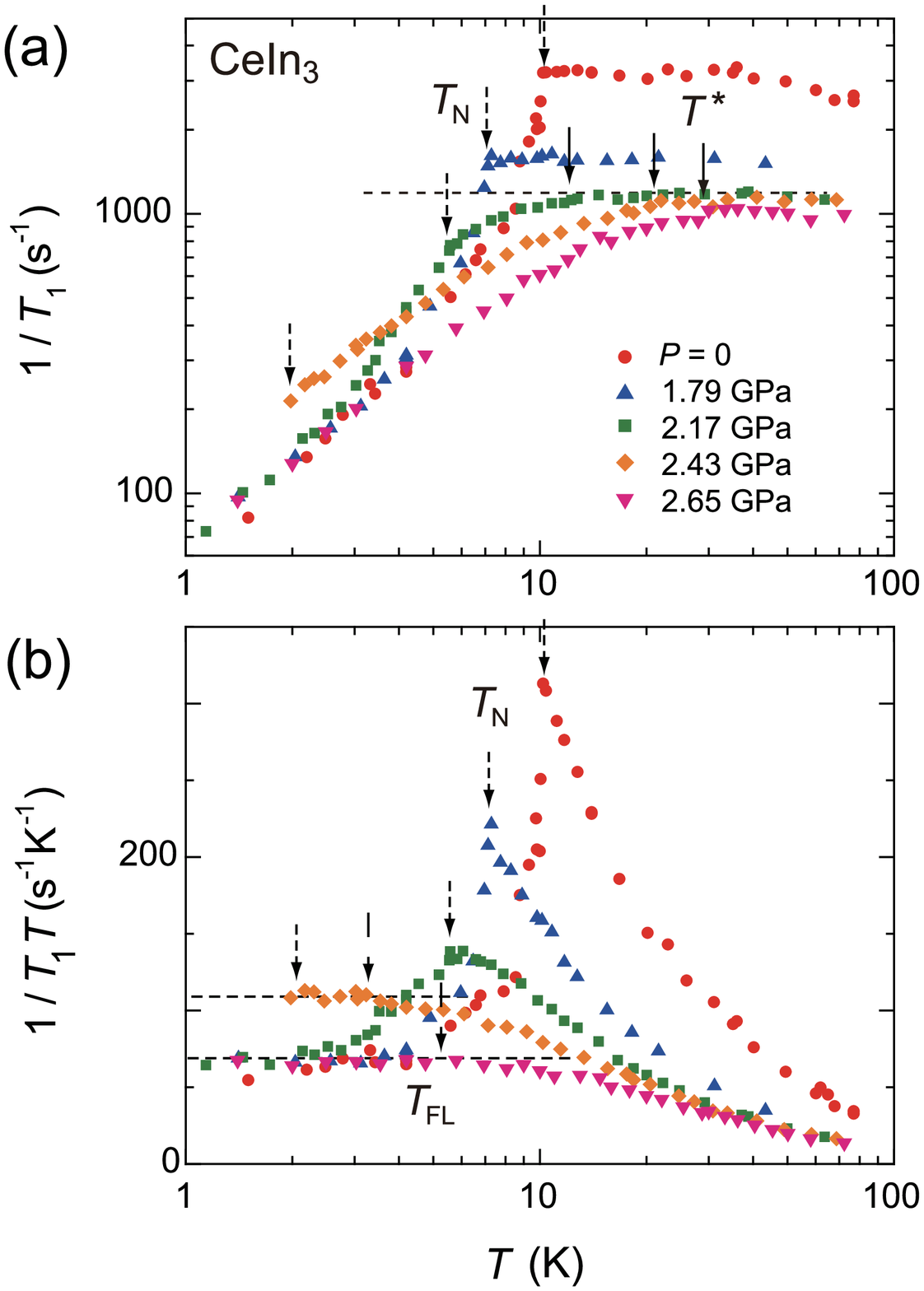}
\caption[]{(color online) (a) Temperature dependence of $1/T_1$ at $P$ = 0, 1.79, 2.17, 2.43, and 2.65 GPa. Dotted line indicates a relation of $1/T_1$ = constant. Dotted and solid arrows point to $T_{\rm N}$ and $T^{\rm *}$, respectively. (b) Temperature dependence of  $1/T_1T$ at $P$ = 0, 1.79, 2.17, 2.43, and 2.65 GPa. Dotted line indicates the relation of $1/T_1T$ = constant. Dotted and dashed arrows indicate $T_{\rm N}$ and $T_{\rm FL}$, respectively.}
\label{fig10}
\end{figure}

Figure~\ref{fig10} shows the $T$ dependences of $1/T_1$ (Fig.~\ref{fig10}(a)) and $1/T_1T$ (Fig.~\ref{fig10}(b)) in CeIn$_3$ at $P$ = 0, 1.79, 2.17, 2.43, and 2.65 GPa. The $1/T_1$ result at $P=0$ is consistent with the previous one \cite{Kohori}. $1/T_1$ at $P$ = 0 shows a gradual increase upon cooling and stays constant till $T_{\rm N}$ below 40 K, which evidences a localized nature of Ce-$4f$ derived  magnetic fluctuations coupled with each other via the Ruderman-Kittel-Kasuya-Yosida (RKKY) interaction.  Below $T_{\rm N}$, $1/T_1$ decreases significantly without any critical slowing down behavior near $T_{\rm N}$. Similar behaviors were reported in $1/T_1$ in CePd$_2$Si$_2$ \cite{YKawasaki}. In this compound, when noting that a Kondo temperature $T_{\rm K}$ is nearly the same as $T_{\rm N}$, the absence of critical magnetic fluctuations towards $T_{\rm N}$ may be relevant to a competition between the Kondo local interaction and the RKKY intersite interaction. Notably, $1/T_1T$ remains constant well below $T_{\rm N}$ at low temperatures, probing a residual Fermi surface in the AFM state.  

\begin{figure}[h]
\centering
\includegraphics[width=6.5cm]{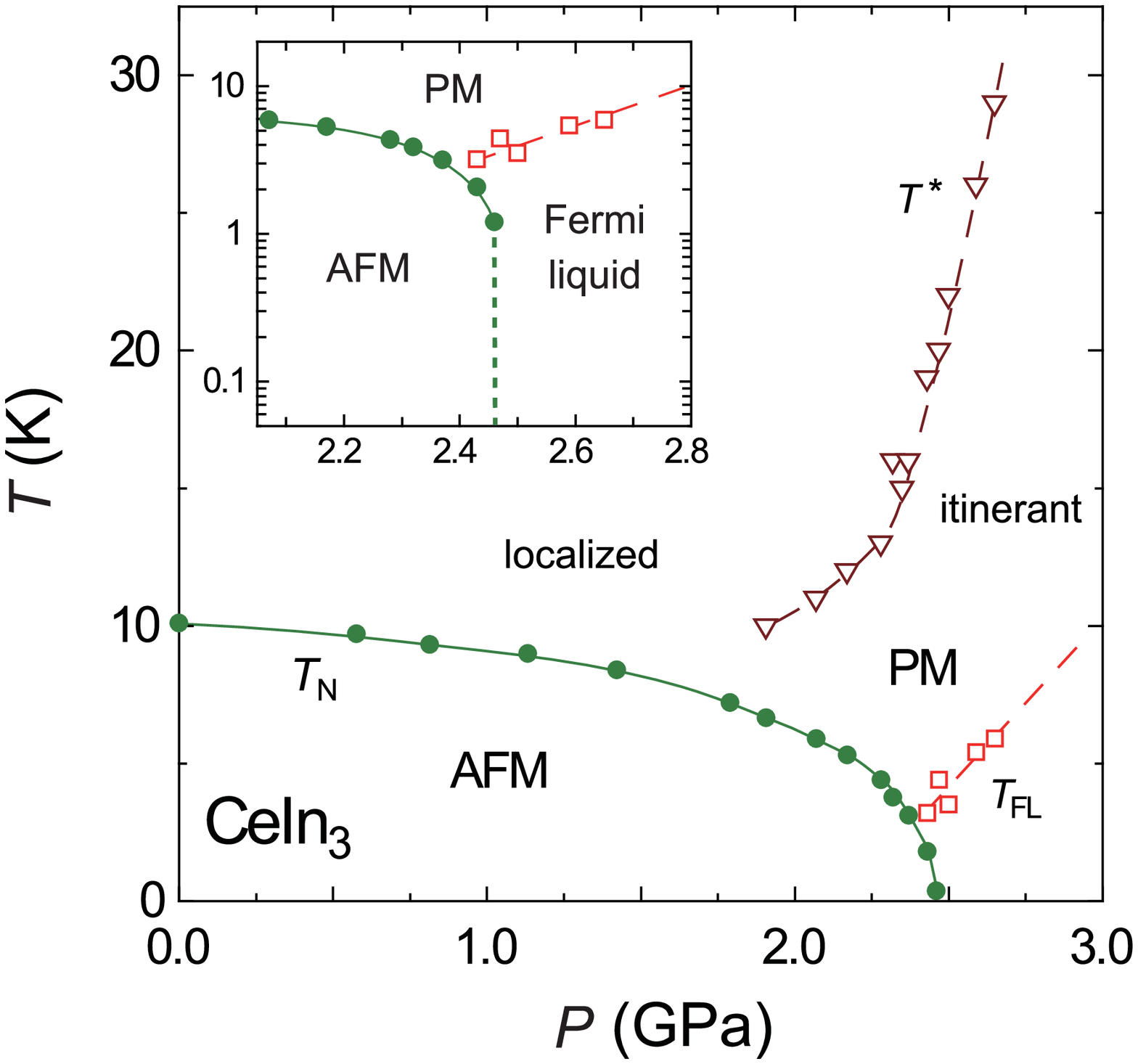}
\caption[]{(color online) The $P-T$ phase diagram of the magnetic properties for CeIn$_3$ obtained from the present NQR experiments. Solid circles indicate $T_{\rm N}$, while open triangles denote a crossover temperature $T^*$ from a localized to an itinerant regime of 4$f$-electrons; open squares indicate a Fermi temperature $T_{\rm FL}$ below which a HF state is established. The inset shows the detailed phase diagram in the vicinity of $P_{\rm c}$ in a semi-logarithmic scale. A vertical dotted line indicates the first-order phase boundary with a minimum value of $T_{\rm N}$ = 1.2 K between the AFM and PM.}
\label{fig11}
\end{figure}

Next, we deal with the results under $P$. Since $1/T_1$ still remains constant just above $T_{\rm N}$ up to $P$ = 1.79 GPa, a localized magnetic character is robust against the application of $P$ in CeIn$_3$. In a localized regime, $1/T_1$ is proportional to $p_{\rm eff}^2/J_{\rm ex}$ or $\sim p_{\rm eff}^2 W/J_{\rm cf}^2$. Here $p_{\rm eff}$, $J_{\rm ex}$, $J_{\rm cf}$, and $W$ are an effective PM local moment, the RKKY exchange constant, the exchange constant between 4$f$ moments and conduction-electron spins, and the bandwidth of conduction electrons, respectively. A progressive suppression of the value of $1/T_1$ = constant at high temperatures with increasing $P$ is considered to be due to a reduction in $p_{\rm eff}$ and/or an increase in $J_{\rm cf}$. As a result, $1/T_1$ starts to decrease below $T^*$ = 10 K in a $P$ range exceeding $P$ $\sim$ 1.9 GPa as observed in Fig.\ref{fig10}(a). In HF systems, it is known that $T^*$ is scaled to the quasi-elastic linewidth in a neutron-scattering spectrum, leading to a tentative estimation of the bandwidth of the HF state. As shown in Fig.\ref{fig11}, as $P$ exceeds $P_{\rm c}$, $T^*$ increases steeply up to $T^*$ $\sim$ 30 K at $P$ = 2.65 GPa \cite{Skawasaki2}.

Figure \ref{fig10}(b) indicates the $P$- and $T$-dependences of $1/T_1T$ that probes low-lying excitations in an itinerant regime. Notably, a behavior that $1/T_1T$ = const. is observed below $T_{\rm FL}$ = 3.2 K at $P$ = 2.43 GPa just below $P_{\rm c}$ = 2.46 GPa. Here, we defined $T_{\rm FL}$ as a Fermi temperature below which $1/T_1T$ becomes constant. As shown in the inset of Fig.\ref{fig11}, the HF state is realized below $T_{\rm FL}$ = 3.2 K at $P$ = 2.43 GPa, which is just below $P_{\rm c}$. Remarkably, $T_{\rm N}\sim 1.2$ K is lower than $T_{\rm FL}$ $\sim$ 4.5 K at $P_{\rm c}$. The HF-PM state and the AFM state compete to trigger the weakly first-order QPT  at $P_{\rm c}$ = 2.46 GPa in the case of CeIn$_3$. The $P$ dependence of $T_{\rm FL}$ is in good agreement with the observation of $T^2$ dependence in resistivity measurements \cite{Knebel}. Non-Fermi liquid behaviors due to the development of AFM spin fluctuations are not evident; one reason is that the HF state is already realized in the PM state above $T_{\rm N}$ just below $P_{\rm c}$, and subsequently, a first-order QPT occurs near $P_{\rm c}$. 

Even though $T_{\rm N}$ is markedly decreased in the vicinity of $P_{\rm c}$, being lower than $T_{\rm FL}$ above $P$ = 2.43 GPa, which suggests an itinerant-type AFM order, we have suggested that as far as AFM order survives, a localized character is robust against closely approaching $P_{\rm c}$. This is because  the $T$ dependences of NQR spectra and their spectral intensity near $P_{\rm c}$ are consistently simulated by assuming a localized character for AFM order. This robustness of the localized nature of AFM order just below $P_{\rm c}$ may be relevant with the first-order nature of QPT taking place at $P_{\rm c}$.

\begin{figure}[h]
\centering
\includegraphics[width=6cm]{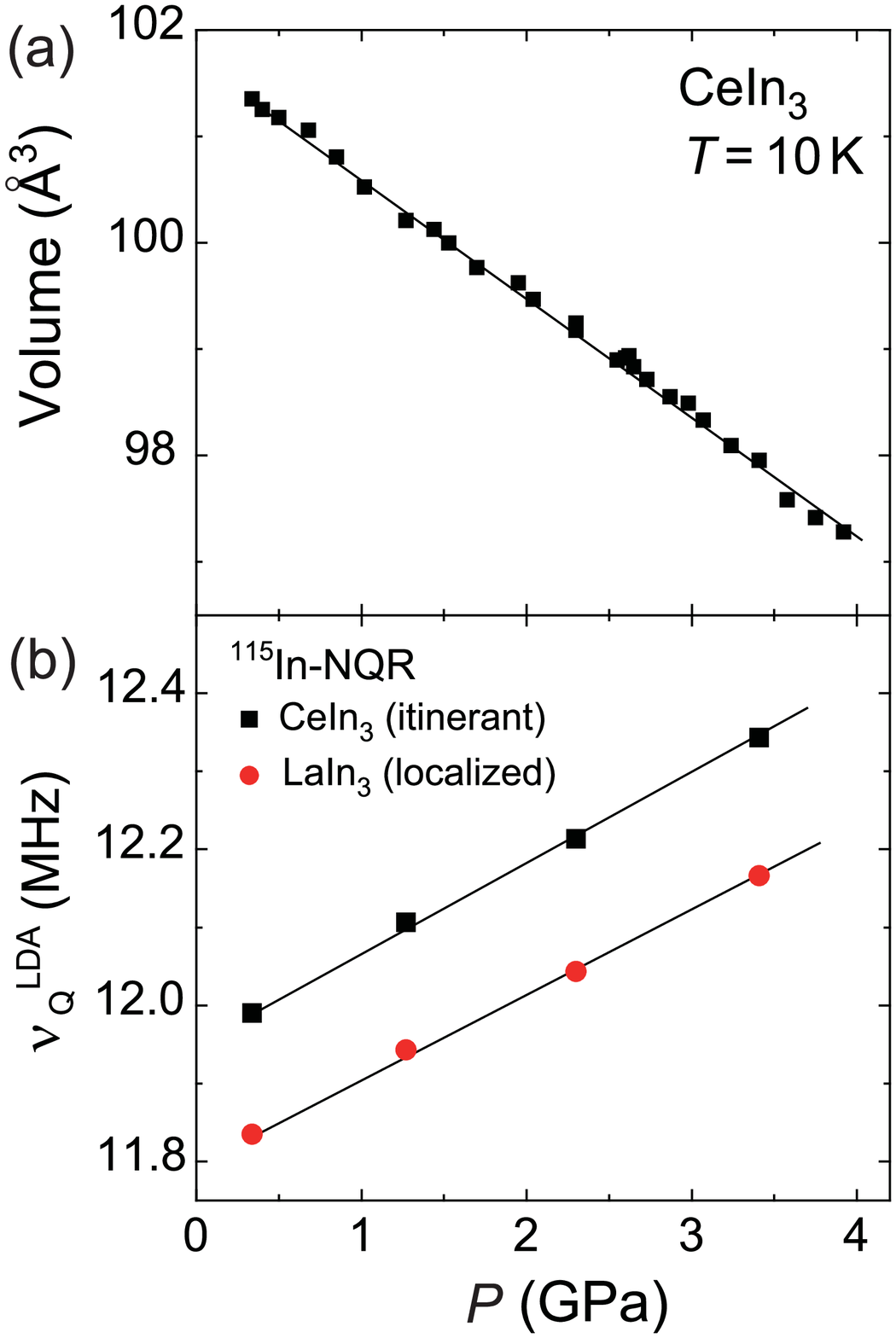}
\caption[]{(color online) (a) Pressure dependence of the lattice volume for CeIn$_3$. (b)  $P$ dependence of $\nu_{\rm Q}$ obtained from the LDA band calculation for CeIn$_3$ and LaIn$_3$. LaIn$_3$ corresponds to the 4$f$ localized model of CeIn$_3$. Both are calculated with the same lattice constant.}
\label{fig12}
\end{figure}

\begin{figure}[h]
\centering
\includegraphics[width=6cm]{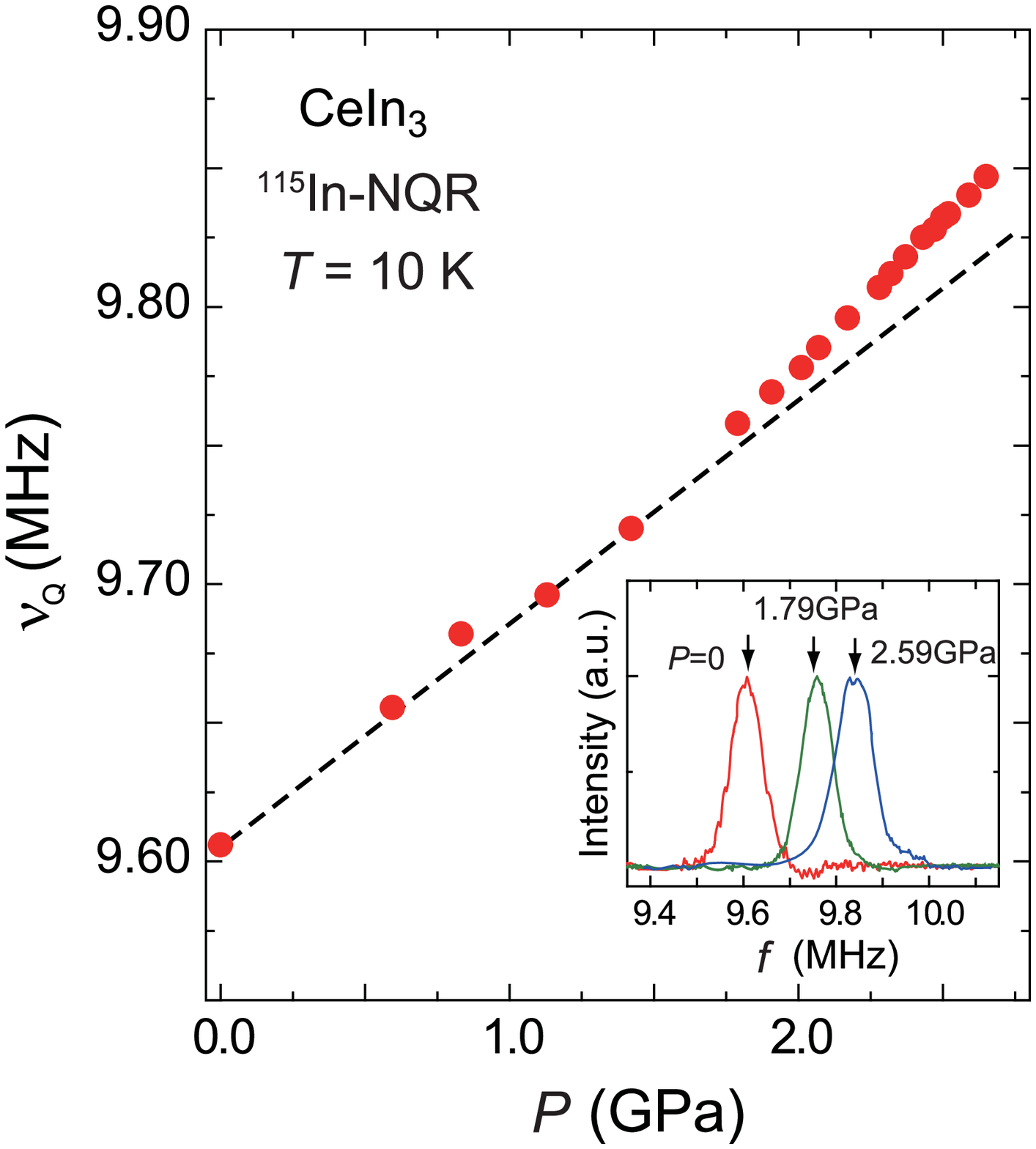}
\caption[]{(color online) Pressure dependence of $\nu_{\rm Q}$ at $T$ = 10 K. Dotted line is an eye-guide. The inset shows 1$\nu_{\rm Q}$-NQR spectra at $P$ = 0, 1.79 and 2.59 GPa. }
\label{fig13}
\end{figure} 

\section{Electronic state behind the first-order transition}

Here, we characterize the first-order magnetic phase transition from the AFM to PM. An NQR frequency $\nu_{\rm Q}$ probes the electric-field gradient (EFG) generated by the electron distribution surrounding the In site. In general, a compression of the lattice volume $V$ increases $\nu_{\rm Q}$; i.e., $\nu_{\rm Q}$ is proportional to $1/V$. In addition to this lattice contribution of EFG, a local electronic charge distribution at the In site generates an electronic contribution of EFG. While the lattice volume of CeIn$_3$ is actually compressed as $P$ increases, as shown in Fig.~\ref{fig12}(a), $\nu_{\rm Q}$ estimated from the LDA band calculation increases for CeIn$_3$; an identical trend is obtained for LaIn$_3$, as shown in Fig.~\ref{fig12}(b). Note that $\nu_{\rm Q}$ for CeIn$_3$ is larger than that for LaIn$_3$. This is considered to be due to the electronic contribution of EFG for CeIn$_3$ that originates from the hybridization between Ce derived 4$f$-electrons and $p$-electrons at the In site. Figure~\ref{fig13} shows the $P$ dependence of the NQR frequency, and the $\nu_{\rm Q}$ value at $T$ = 10 K. Here, the NQR spectra with $P$, which are displayed in the inset of Fig.~\ref{fig13}, are obtained by the Fourier-transform method of the spin-echo signal. As observed in Fig.~\ref{fig13}, in a lower $P$ region, the compression of the lattice volume increases $\nu_{\rm Q}$ linearly. As $P$ increases beyond 2 GPa where $T^*$ starts to increase, $\nu_{\rm Q}$ begins to increase significantly. The origin of a larger increasing rate of $\nu_{\rm Q}$ at pressures larger than 2 GPa is due to the increase in the electric contribution that is related to the significant increase in hybridization between $f$-electrons and conduction electrons. As a result, $T_{\rm N}$ decreases steeply with the increase in $P$. Notably, such a variation from a localized to an itinerant nature of the $f$-electrons around $P_{\rm c}$ is also reported by the recent dHvA measurement under $P$ \cite{Settai}.
   
\section{Pressure-induced superconductivity around $P_{\rm c}$}

\begin{figure}[h]
\centering
\includegraphics[width=6cm]{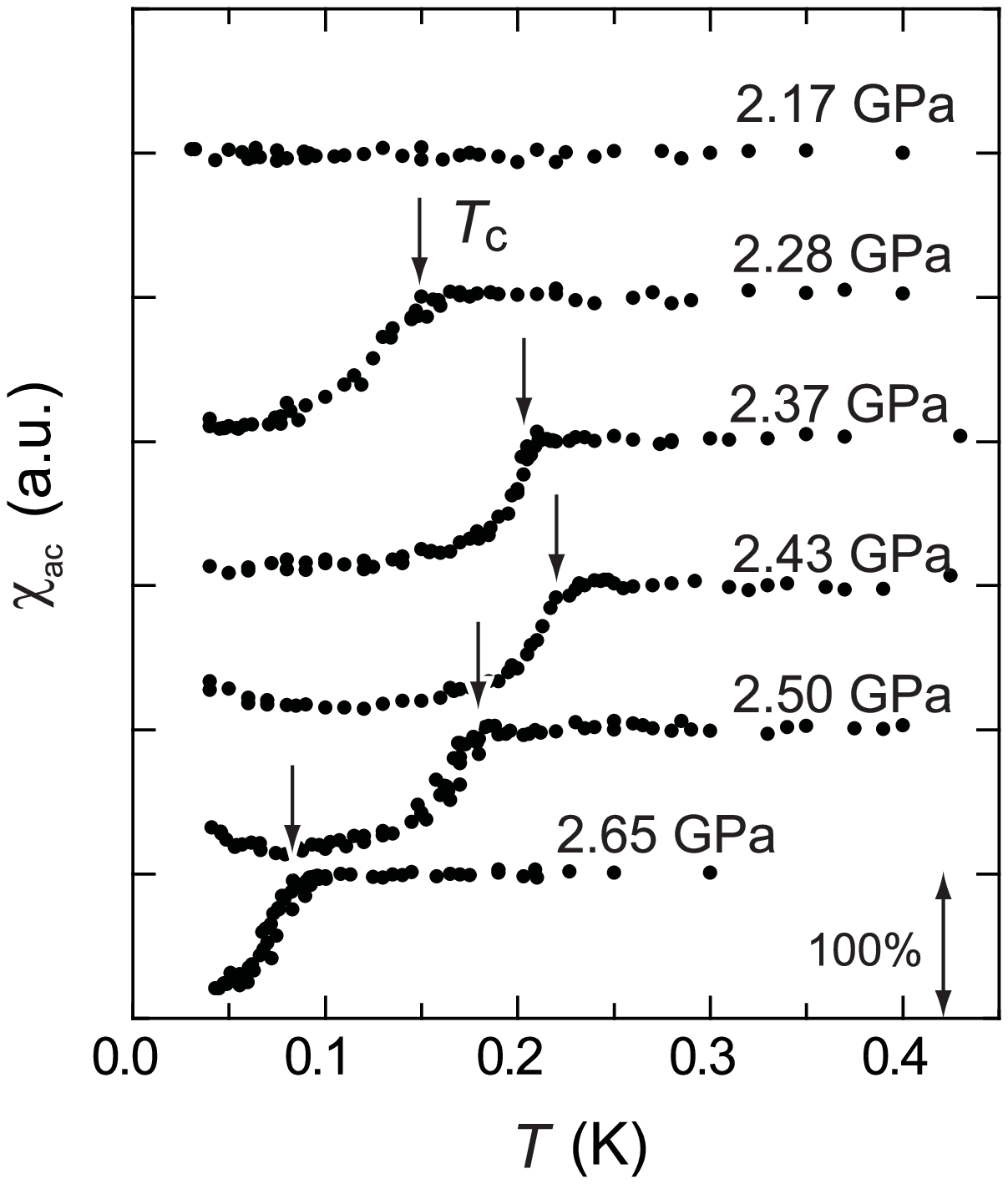}
\caption[]{(color online) Temperature dependence of $\chi_{\rm ac}$ in a $P$ range of  $2.17 - 2.65$ GPa. Arrows indicate an onset temperature $T_{\rm c}^{\rm onset}$ below which SC diamagnetism starts to appear.}
\label{fig14}
\end{figure}

\begin{figure}[h]
\centering
\includegraphics[width=6.7cm]{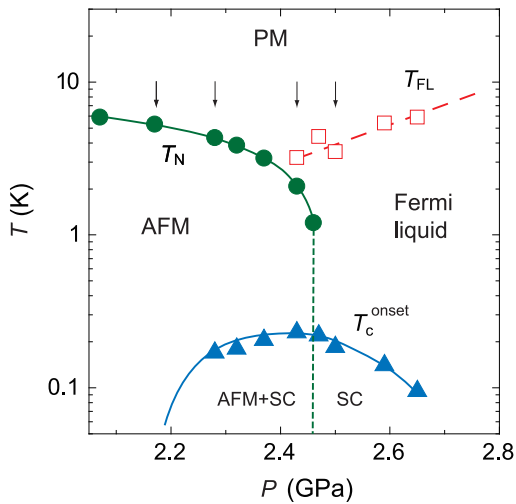}
\caption[]{(color online) Detailed $P-T$ phase diagram for PM, AFM, and SC for CeIn$_3$ in the vicinity of the first-order QPT at $P_{\rm c}$ = 2.46 GPa, which is denoted by a vertical dashed line. Solid circles and triangles indicate $T_{\rm N}$ and $T_{\rm c}$, respectively. Open square is a crossover temperature towards the HF state. $T_{\rm c}^{\rm onset}$ is determined by $\chi_{\rm ac}$ in a $P$ range of $2.17 - 2.65$ GPa. Arrows point to $P$ = 2.17, 2.28, 2.43, and 2.50 GPa where the $1/T_1$ shown in Fig.~\ref{fig16} was measured (see text).}
\label{fig15}
\end{figure}

In this section, the $P$-induced superconductivity in CeIn$_3$ is considered.
Figure~\ref{fig14} shows the $T$ dependence of the ac-susceptibility $\chi_{\rm ac}$ of CeIn$_3$ measured by an $in-situ$ NQR coil under $P$. In the range $P$ = 2.28 - 2.65 GPa, a clear decrease in $\chi_{\rm ac}$ implies a SC transition under $P$. However, the SC does not occur at $P$ = 2.17 GPa till $T$ = 30 mK, as can be noticed from the $T$ independence of $\chi_{\rm ac}$ shown in Fig.\ref{fig14}. Thus, a critical pressure $P_{\rm SC}$ for the onset of SC is between $P$ = 2.17 and 2.28 GPa. The absolute value of $\chi_{\rm ac}$ was corrected by a value of $\chi_{\rm ac}$ measured on a single crystal of CeIrIn$_5$ in which the bulk SC is fully established \cite{Sumiyama}.  Figure \ref{fig15} shows a SC phase diagram as the function of $P$. The $P$-induced SC emerges in the vicinity around $P_{\rm c}$ where the first-order transition occurs. This SC phase for CeIn$_3$ is consistent with other experiments \cite{Mathur,Muramatsu,Knebel,Settai}. It must be noted that the SC volume fraction is almost unchanged in the range $P$ = 2.28 - 2.65 GPa. Thus, the coexistence of the AFM and SC is strongly indicative of the $P$ range of $P = 2.28$ GPa $-$ $P_{\rm c}$ = 2.46 GPa.

\section{Evidence for the uniformly coexisting phase of antiferromagnetism and unconventional superconductivity}

\begin{figure}[h]
\centering
\includegraphics[width=7cm]{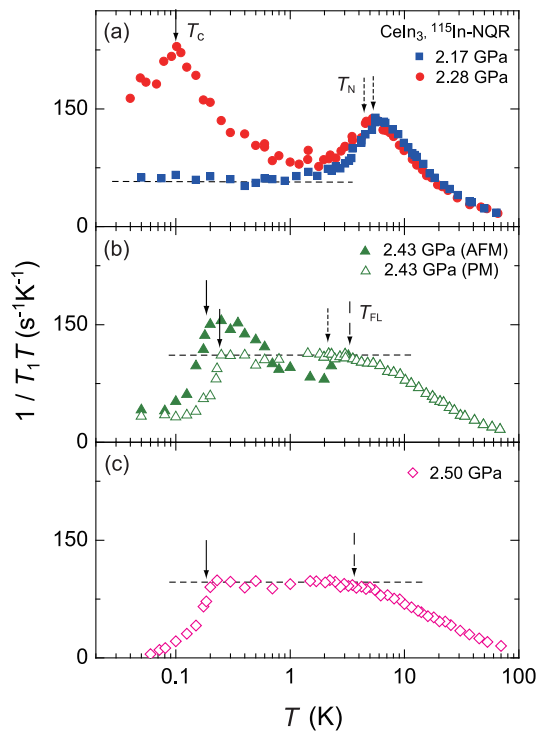}
\caption[]{(color online) Temperature dependence of $^{115}(1/T_1T)$ for CeIn$_3$ at (a) $P$ = 2.17 and 2.28 GPa, (b) 2.43 GPa, and (c) 2.50 GPa. Open and solid symbols indicate the respective data for PM and AFM (see the text). The solid arrows indicate the respective SC transition temperature $T_{\rm c}^{\rm PM}$ and $T_{\rm c}^{\rm AFM}$ for PM and AFM. The dotted and dashed arrows indicate $T_{\rm N}$ and $T_{\rm FL}$ below which the HF state becomes valid, characterized by the $T_1T$=const.law (dotted line).}
\label{fig16}
\end{figure}

The coexistence of the AFM and SC is directly evidenced from the $T$ dependence of $1/T_1T$ that can probe low-lying excitations due to the quasiparticles in SC and also due to magnetic excitations in AFM.  Figures~\ref{fig16} (a), (b), and (c) show drastic changes in the $T$ dependence of $1/T_1T$ at (a) $P$ = 2.17 and 2.28, (b) $P$ = 2.43 GPa, and (c) $P$ = 2.50 GPa, respectively. Each pressure is indicated by the arrow in Fig. \ref{fig15}.  Note that since $P$ = 2.43 GPa is very close to $P_{\rm c}$ = 2.46 GPa where the first-order transition occurs, the $1/T_1T$'s for PM and AFM are separately measured on the respective NQR spectral peaks (see the middle spectrum in Fig. \ref{fig6}), as shown in Fig. \ref{fig16}(b). Here, $T_{\rm c}$ is determined as a temperature below which $1/T_1T$ decreases significantly due to the SC gap opening, and the NQR intensity begins to decrease due to the Meissner shielding of rf pulses. Since the coherence peak is absent just below $T_{\rm c}$, it suggests that unconventional superconductivity is induced in CeIn$_3$ under pressure. These results give microscopic evidence for the uniformly coexisting phase of the AFM and SC in the range $P = 2.28 - 2.43$ GPa for CeIn$_3$. We note here that the behavior of $1/T_1T$ = const. observed at $P$ = 2.37 GPa (see Fig.19(a)) reveals that the SC phase in the AFM+SC uniformly coexisting state does not exhibit a line-node gap, but is in a gapless regime. This is in contrast with the SC phase in the PM phase at $P$ = 2.50 GPa where the $T^3$ dependence of $1/T_1$ is observed as discussed later.

\begin{figure}[h]
\centering
\includegraphics[width=6.5cm]{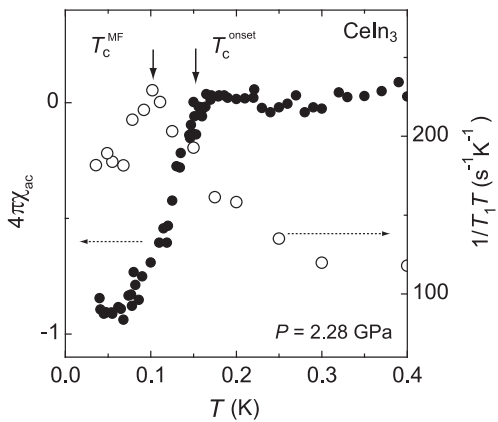}
\caption[]{ Temperature dependence of $^{115}(1/T_1T)$ (open circle) and $\chi_{\rm ac}$ (solid circle) for CeIn$_3$ at $P$ = 2.28 GPa where the SC emerges. Arrows point to $T_{\rm c}^{\rm MF}$ below which $^{115}(1/T_1T)$ decreases and $T_{\rm c}^{\rm onset}$ below which the SC diamagnetism appears.}
\label{fig17}
\end{figure}

Figure~\ref{fig17} demonstrates the SC characteristics for the coexistence of AFM and SC at $P$ = 2.28 GPa. Below $T$ = 3 K, $1/T_1T$ continues to increase moderately down to $T_{\rm c}^{\rm MF}$ = 0.1 K even though it crosses $T_{\rm c}^{\rm onset}\sim 0.15$ K. This relaxation behavior suggests that the SC order parameter does not always develop below $T_{\rm c}^{\rm onset}$. These characteristics for the uniformly coexisting phase of the AFM and SC are similar to those for the $P$-induced superconductor CeRhIn$_5$ \cite{Skawasaki,Yashima}. Noting that $T_{\rm c}^{\rm MF}$ coincides with $T_{\rm c}^{\rm onset}$ at $P$ = 2.65 GPa with $T_{\rm c}$ = 95 mK \cite{Shinji}, the difference  between $T_{\rm c}^{\rm MF}$ and $T_{\rm c}^{\rm onset}$ at $P$ = 2.28 GPa is not due to the $P$ distribution but due to the uniform coexistence of the AFM and SC. 

Next, we focus on the novel low-lying magnetic excitations inside AFM.
As shown in Fig.\ref{fig16}(a), the $1/T_1T=$const. behavior is observed well below $T_{\rm N}$ = 5.3 K at $P$ = 2.17 GPa as well as at $P$ = 0\cite{Kohori,Kohori2}. Unexpectedly, $1/T_1T$ at $P=2.28$ GPa, where $T_{\rm N}$ starts to steeply decrease, continues to increase upon cooling below $T_{\rm N}$ and exceeds the value at $T_{\rm N}$= 4.4 K regardless of AFM spin polarization being induced. This behavior is also observed in the AFM state at $P$ = 2.37 and 2.43 GPa, as shown in Fig.~\ref{fig16}(b) and Fig.~\ref{fig19}(a), respectively. Since $1/T_1T$ probes any type of low-lying excitations, it is probable that low-lying longitudinal spin-density fluctuations are responsible for this feature in association with the first-order QPT. This is because when the AFM critical temperature is suppressed at the termination point of the first-order QPT, i.e., when $P_{\rm c}=2.46$ GPa, the diverging AFM spin-density fluctuations emerge at a critical point from AFM to PM. 
Namely, since a free energy of the system in the vicinity of $P_{\rm c}$ becomes almost the same between the AFM phase with a finite spin polarization at frequencies ($\omega\sim$0) lower than an NMR frequency and the PM-HF state which does not carry static spin polarization but is dominated by low-lying excitations, it is likely that an amplitude of spin-density is fluctuating in the vicinity of $P_{\rm c}$.
In this context, the $P$-induced SC in CeIn$_3$ does not always occur with the background of a magnetically soft-electron liquid state \cite{Mathur}, but instead, novel magnetic excitations, such as AFM spin-density fluctuations relevant to a first-order transition from the AFM to PM might mediate attractive interaction. Irrespective of the pairing mechanism is at $P$ = 2.28 GPa where the AFM order is realized over the entire sample below $T$ = 3 K (see Fig. \ref{fig7}), the clear decrease in $1/T_1T$ and $\chi_{ac}$ provides convincing evidence for the uniformly coexisting phase of the AFM and SC in CeIn$_3$. These results suggest that the $P$-induced SC in the AFM phase is closely related to the enhancement of hybridization that triggers the QPT from the AFM to PM.

Significantly, the SC coexisting with AFM in CeIn$_3$ is reasonably unique as expected from the results at $P$ = 2.43 GPa, as shown in Fig. \ref{fig16}(b). At temperatures lower than the respective values of $T_{\rm c}^{\rm PM}$ = 230 mK and $T_{\rm c}^{\rm AFM}$ = 190 mK for PM and AFM phases, unexpectedly, the magnitudes of $1/T_1T$ =  const. coincide with one another; nevertheless, both phases are separated across $P_{\rm c}$ where the first-order QPT occurs and the values of $T_{\rm c}$ differ. This means that the low-lying excitations may be identical in origin for the uniformly coexisting state of AFM+SC and for the PM+SC. How does this happen? It may be possible that both phases are in a dynamically separated regime with time scales smaller than the inverse of NQR frequency in order to make each SC phase for AFM and PM uniform across $P_{\rm c}$. In this context, the observed magnetically separated phases and the relevant SC coexisting with the AFM may belong to new phases of matter. 
\begin{figure}[h]
\centering
\includegraphics[width=6.5cm]{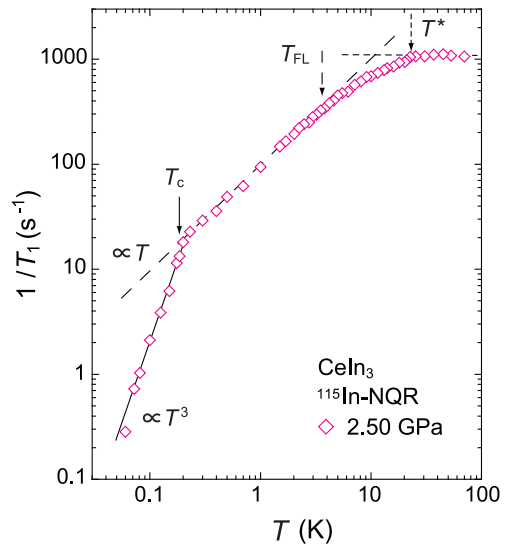}
\caption[]{(color online) Temperature dependence of $1/T_1$ at $P=2.50$ GPa just above $P_{\rm c}$ = 2.46 GPa. Solid, dashed, and dotted arrows indicate $T_{\rm c}$, $T_{\rm FL}$, and $T^{*}$, respectively. Solid, dashed, and dotted lines indicate the respective relations of $1/T_1\propto T^3$, $1/T_1\propto T$, and $1/T_1$ = constant.}
\label{fig18}
\end{figure}

\section{Heavy-fermion superconductivity}
\begin{figure}[h]
\centering
\includegraphics[width=8.6cm]{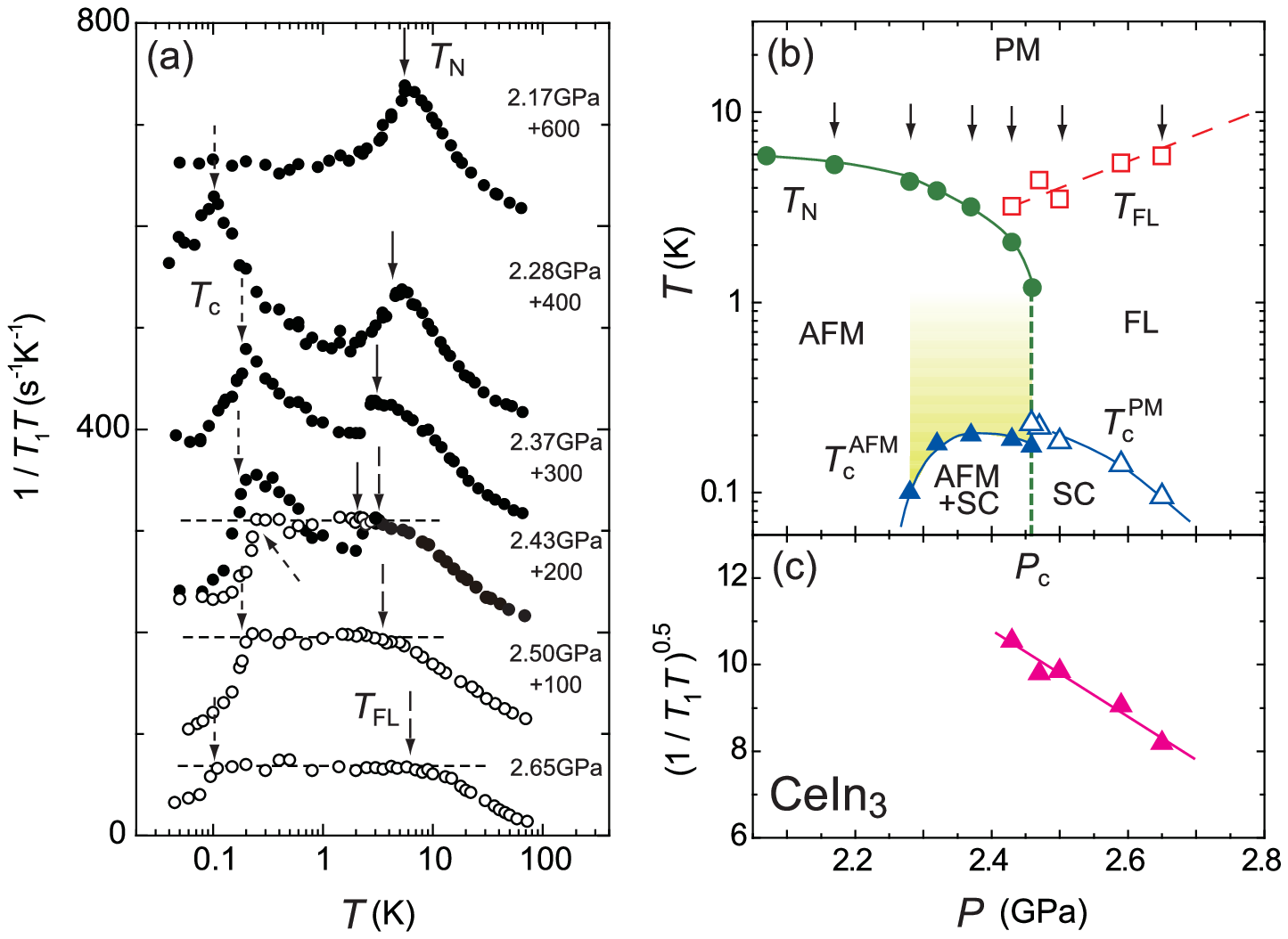}
\caption[]{(color online) (a) Temperature dependences of $1/T_1T$ in a range $P= 2.17-2.65$ GPa. The $1/T_1T$ data are offset for clarity. The solid and open symbols indicate the respective data of $1/T_1T$ measured below and above $P_{\rm c}$. Solid, dotted, and dashed arrows indicate $T_{\rm N}$, $T_{\rm c}$, and $T_{\rm FL}$, respectively. Dotted lines indicate a relation of $1/T_1T$ = constant. (b) The detailed $P-T$ phase diagram of CeIn$_3$ in the vicinity of $P_{\rm c}$.  Shaded region indicates the unconventional magnetic state where the low-lying spin-density fluctuations develop down to $T_{\rm c}$ upon cooling. All these phases of matter are determined by the present NQR measurements under $P$ (see text). Arrow points to a value of $P$ where $1/T_1T$ in (a) were measured. (c) Pressure dependence of $(1/T_1T)^{\rm 0.5}$ for $P_{\rm c} < P$ which is in proportion to the effective density of states at the Fermi level of HF band. Solid line is an eye-guide.}
\label{fig19}
\end{figure}

Finally, we present evidence for the $P$-induced unconventional HF SC emerging at the PM phase beyond $P_{\rm c}$  where the low-lying AFM spin fluctuations are absent. This is because, as observed in Fig. \ref{fig18}, $1/T_1T$ = const. behavior is obeyed at the normal state, and $1/T_1$ follows a $T^3$ dependence below $T_{\rm c}$ = 185 mK without the coherence peak just below $T_{\rm c}$; this is consistent with the line-node gap model characteristic for an unconventional HF SC \cite{KitaokaKuramoto,Ishida,Mito2,Zheng,Ykawasaki3,Ykawasaki2,Shinji,Shinji2,Yashima2,Kitaoka}. As observed in Fig.~\ref{fig19}(a), the respective Fermi-liquid temperature $T_{\rm FL}$ at $P$ = 2.50 GPa and $P$ = 2.65 GPa are defined as $T_{\rm FL}$ = 3.5 K and 5.9 K, respectively; below this temperature, the $1/T_1T$ = const. behavior is valid. Noting that the value of $(1/T_1T)^{1/2}$ is proportional to the effective density of state at the Fermi level, the value of $(1/T_1T)^{1/2}$ has a maximum at $P_{\rm c}$ where $T_{\rm c}$ has the maximum value of $T_{\rm c}^{\rm max}$ = 230 mK as observed in Fig.~\ref{fig19}(c).
This means that as an effective HF bandwidth becomes smaller and the system approaches $P_{\rm c}$,  $T_{\rm c}$ is increased up to the maximum value. This result reveals that the $P$-induced SC in CeIn$_3$ is realized under a strong electron correlation, although the antiferromagnetic QCP is absent. In this context, the first-order QPT plays an important role for the onset of unconventional HF SC as well.  

\section{Conclusion}
The extensive $^{115}$In-NQR studies conducted under $P$ on CeIn$_3$ have revealed the evolution of magnetic properties and $P$-induced  unconventional SC characteristics as follows:
\begin{enumerate}
\item The $P$-induced transition from an AFM to PM is the first-order QPT at a critical pressure $P_{\rm c}$ = 2.46 GPa at which the AFM order disappears with a minimum value of $T_{\rm N}(P_{\rm c})$ = 1.2 K. 
\item The hybridization between $4f$ electrons and conduction electrons increases beyond $P$ = 2 GPa, thereby stabilizing the HF-PM state. It is this competition between the AFM phase, where $T_{\rm N}$ is reduced, and the formation of the HF-PM phase that triggers the first-order QPT at $P_{\rm c} = 2.46$ GPa. 
\item Despite the lack of an AFM QCP in the $P-T$ phase diagram, the unconventional SC occurs in both phases of the AFM and PM. As a result, the AFM order uniformly coexists with the SC order. 
\item The significant increase in $1/T_1$ upon cooling in the AFM phase has revealed the development of low-lying magnetic excitations till $T_{\rm c}$, and this is related to the onset of the uniformly coexisting phase of SC+AFM. 
\item In the HF-PM phase where AFM spin fluctuations are absent, $1/T_1$ decreases without the coherence peak just below $T_{\rm c}$, followed by a power-law like $T$ dependence that indicates an unconventional SC with a line-node gap. 
\item $T_{\rm c}$ has a peak around $P_{\rm c}$ in the HF-paramagnetic phase as well as in the AFM phase and an SC dome exists with a maximum value of $T_{\rm c}$ = 230 mK around $P_{\rm c}$. These results suggest that an origin for the $P$-induced HF SC in CeIn$_3$ is {\it not related to the AFM spin fluctuations but is related to the emergence of the first-order QPT} at $P_{\rm c}$ = 2.46 GPa. 

\end{enumerate}

These novel phenomena observed in CeIn$_3$ should be understood in terms of the first-order QPT because these new phases of matter are induced by applying $P$. When the AFM critical temperature is suppressed at the termination point of the first-order QPT, i.e., when $P_{\rm c}$ = 2.46 GPa, it is anticipated that the diverging AFM spin-density fluctuations emerge at the critical point from AFM to PM. The results on CeIn$_3$ leading to a new type of quantum criticality deserve further theoretical investigations.

\section*{Acknowledgment}
S. K. thanks Yuki Fuseya and Shinji Watanabe for their valuable discussions, comments, and encouragement. S. K. also thanks T. Mito, A. V. Kornilov, C. Thessieu, Y. Kawasaki, H. Kotegawa, K. Ishida, T. Muramatsu, J. Flouquet, and G.-q. Zheng for their assistance during the experiments and/or useful discussions during the early stage of this work. This work was supported by a Grant-in-Aid for Creative Scientific Researchi15GS0213), MEXT and The 21st Century COE Program supported by the Japan Society for the Promotion of Science. The synchrotron radiation experiments were performed at BL10XU in SPring-8 with the approval of the Japan Synchrotron Radiation Research Institute (Proposal No.2001A0004-LD -np).

*Present address: Department of Physics, Faculty of Science, Okayama University, Okayama 700-8530, Japan
**Present address: Department of Physics, Graduate School of Science, Kyoto University, Kyoto 606-8502, Japan


\end{document}